\documentclass[aps, reprint, superscriptaddress,prl]{revtex4-2}
\usepackage{siunitx}
\usepackage{graphicx}
\usepackage{placeins}
\usepackage{textcomp}
\usepackage{amssymb}
\usepackage{amsmath}
\usepackage{times, txfonts}
\usepackage{makecell}
\usepackage{physics}
\usepackage[english]{babel}
\usepackage[colorlinks=true,linkcolor=blue,urlcolor=blue,citecolor=blue,pdfusetitle]{hyperref}

\begin{document}
\title{Realisation of a Coherent and Efficient One-Dimensional Atom}
\author{Natasha Tomm}
\thanks{These authors contributed equally}
\affiliation{Department of Physics, University of Basel, Klingelbergstrasse 82, CH-4056 Basel, Switzerland}
\author{Nadia O. Antoniadis}
\thanks{These authors contributed equally}
\affiliation{Department of Physics, University of Basel, Klingelbergstrasse 82, CH-4056 Basel, Switzerland}
\author{Marcelo Janovitch}
\thanks{These authors contributed equally}
\affiliation{Department of Physics, University of Basel, Klingelbergstrasse 82, CH-4056 Basel, Switzerland}
\author{Matteo Brunelli}
\affiliation{Department of Physics, University of Basel, Klingelbergstrasse 82, CH-4056 Basel, Switzerland}
\author{R\"{u}diger Schott} 
\affiliation{Lehrstuhl f\"{u}r Angewandte Festk\"{o}rperphysik, Ruhr-Universit\"{a}t Bochum, D-44780 Bochum, Germany}
\author{Sascha R. Valentin}
\affiliation{Lehrstuhl f\"{u}r Angewandte Festk\"{o}rperphysik, Ruhr-Universit\"{a}t Bochum, D-44780 Bochum, Germany}
\author{Andreas D. Wieck}
\affiliation{Lehrstuhl f\"{u}r Angewandte Festk\"{o}rperphysik, Ruhr-Universit\"{a}t Bochum, D-44780 Bochum, Germany}
\author{Arne Ludwig}
\affiliation{Lehrstuhl f\"{u}r Angewandte Festk\"{o}rperphysik, Ruhr-Universit\"{a}t Bochum, D-44780 Bochum, Germany}
\author{Patrick P. Potts}
\affiliation{Department of Physics, University of Basel, Klingelbergstrasse 82, CH-4056 Basel, Switzerland}
\author{Alisa Javadi} 
\altaffiliation{Present address: School of Electrical and Computer Engineering, Department of Physics and Astronomy, The University of Oklahoma, 110 West Boyd Street, OK 73019, USA} 
\affiliation{Department of Physics, University of Basel, Klingelbergstrasse 82, CH-4056 Basel, Switzerland}
\author{Richard J. Warburton}
\email[To whom correspondence should be addressed:]{\\ m.janovitch@unibas.ch, nadia.antoniadis@unibas.ch}
\affiliation{Department of Physics, University of Basel, Klingelbergstrasse 82, CH-4056 Basel, Switzerland}

\date{\today}

\begin{abstract}
A quantum emitter interacting with photons in a single optical-mode constitutes a one-dimensional atom. A coherent and efficiently coupled one-dimensional atom provides a large nonlinearity, enabling photonic quantum gates. Achieving a high coupling efficiency  ($\beta$-factor) and low dephasing is challenging. Here, we use a semiconductor quantum dot in an open microcavity as an implementation of a one-dimensional atom. With a weak laser input, we achieve an extinction of 99.2\% in transmission and a concomitant bunching in the photon statistics of $g^{(2)}(0) = 587$, showcasing the reflection of the single-photon component and the transmission of the multi-photon components of the coherent input. The tunable nature of the microcavity allows $\beta$ to be adjusted and gives control over the photon statistics -- from strong bunching to anti-bunching -- and the phase of the transmitted photons. We obtain excellent agreement between experiment and theory by going beyond the single-mode Jaynes-Cummings model. Our results pave the way towards the creation of exotic photonic states and two-photon phase gates. 

\end{abstract}
\maketitle

\textit{Introduction.} The ability to generate and manipulate correlated and entangled photonic states at the few-photon level is imperative for the advancement of photon-based quantum technologies. The realisation of quantum photonic gates requires a highly nonlinear medium, i.e., a medium which enables a strong and controlled interaction of few photons \cite{chang_quantum_2014,hartmann_quantum_2016, Chang2018RMP}. A one-dimensional atom, a quantum emitter coupled to a single optical mode, is the ideal candidate to provide these functionalities \cite{shen_strongly_2007}. Engineering a one-dimensional atom is challenging: the photon-emitter coupling efficiency, $\beta$,
should be close to unity, and the emitter should be free of decoherence and noise. One approach is to employ an ensemble of atoms \cite{sorensen_coherent_2016, Prasad2020}, which collectively behaves as a super-atom, or a single emitter in a waveguide for which very high $\beta$-factors have been achieved \cite{arcari_near-unity_2014, Hallett2018, LeJeannic2021PRL, LeJeannic2021}. Cavity quantum electrodynamics provides an alternative route to a high $\beta$-factor: a single emitter is embedded in a microcavity. This approach has been implemented with atoms \cite{goban_atomlight_2014, Tiecke2014Nature, luan_integration_2020}, ions \cite{takahashi_strong_2020}, molecules \cite{wang_turning_2019, pscherer_single-molecule_2021}, and semiconductor quantum dots (QDs) \cite{rakher_externally_2009, de_santis_solid-state_2017, Najer2019}.
    
Here, we embed a single QD in a one-sided microcavity, to create a one-dimensional atom following the original proposal of Ref.~\cite{auffeves-garnier_giant_2007}. Important features are the exceptional coherence (low charge-noise \cite{Kuhlmann2013}, weak dephasing via phonons \cite{IlesSmith2017}) and the high $\beta$-factor. We showcase the performance by measuring the transmission and reflection and their respective $g^{(2)}$-functions. In the ideal case, the emitter acts as a perfect mirror for single photons \cite{shen_coherent_2005, Shen2007, Chang2007}. Our system shows an extinction of 99.2\% of the transmitted light when probed with a low-power laser. Moreover, the transmitted state is highly bunched, $g^{(2)}(0)=587$, a strong demonstration of the nonlinearity at the single-photon level.

The most striking results are the strong extinction and the high bunching of the transmitted state, both metrics for the nonlinearity, both much higher than in previous realisations \cite{Javadi2015NCOM, Hallett2018, wang_turning_2019, pscherer_single-molecule_2021, LeJeannic2021PRL}. Beyond this, first, we exploit the in situ tunability of the microcavity to tailor the photon statistics, transitioning from highly bunched to anti-bunched photonic states. Second, we present a full theoretical model. It includes two optical transitions and two cavity modes thereby going beyond the standard Jaynes-Cummings model. These details are crucial to describe the experimental results fully. Third, we exploit the Rice--Carmichael description of the cavity field \cite{rice_single-atom_1988} to obtain intuitive analytical expressions which explain the main observations.

\begin{figure*}[t!]
    \centering
    \includegraphics{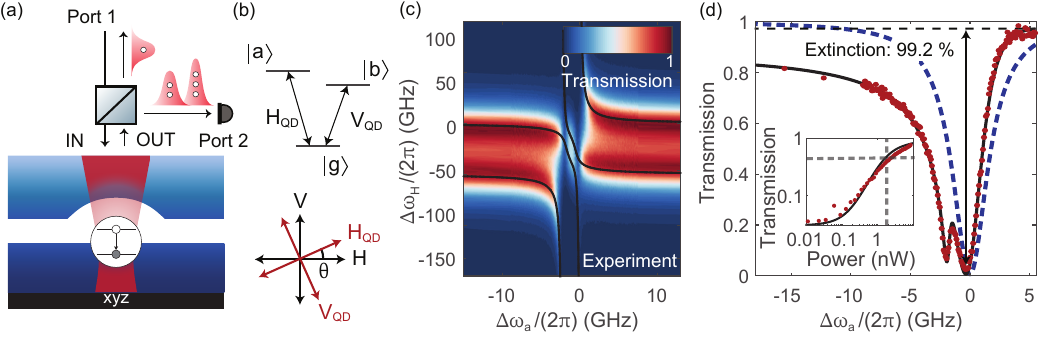} 
    \caption{\textbf{A one-dimensional atom.}
    \textbf{(a)} Experimental setup. Weak laser-light impinges on a one-sided microcavity. The reflected light strikes a polarising beam-splitter. A half-wave plate sets an angle between the input and the axes of the cavity. The sketch shows the path for photons in transmission mode; the input is polarised at 45$^\circ$ relative to the cavity modes. At ideal QD-cavity coupling, single-photon components are reflected into port 1 and multi-photon components enter port 2.
    \textbf{(b)} Top: level structure of the neutral exciton. The two transitions have orthogonal linear polarisations. Bottom: polarisation orientation of the horizontally ($\textrm{H}$) and vertically ($\textrm{V}$) polarised cavity modes and the QD transitions.
    \textbf{(c)} Transmission as a function of cavity detuning ($\Delta\omega_\text{H}$) and QD detuning ($\Delta\omega_a$).
    Electrical tuning was employed keeping the laser frequency fixed: $\Delta\omega_\text{H}$ is tuned via the piezo controlling $z$, the sample--top-mirror separation; $\Delta\omega_a$ is tuned via the voltage applied to the diode.
    Black lines indicate resonances between the laser frequency and the lowest transition frequencies of the Hamiltonian, see \cite{suppmat}.
    \textbf{(d)} Transmission for $\Delta\omega_\text{H}=0$. The transmission features two dips corresponding to the two QD transitions. The stronger transition shows an extinction of 99.2\%. The black line is the full theoretical prediction showing excellent agreement with the data. The dashed blue-line shows the JC model. The inset shows the power dependence of the transmission dip, with $P_{\rm sat}=$1.8 nW (vertical dashed-line).}
    \label{fig:fig1}
\end{figure*}

\textit{The experiment.} The setup is depicted in Fig.\,\ref{fig:fig1}\,(a). The cavity is an open microcavity \cite{Barbour2011} (highly miniaturised Fabry-Perot-type). The bottom mirror is a highly reflective (99.97\%) semiconductor (AlAs/GaAs) DBR mirror. A layer of InAs QDs is embedded within an n-i-p diode structure \cite{Kuhlmann2013}. The top mirror is a less reflective (99\%) dielectric DBR mirror (SiO$_2$/Ta$_2$O$_5$) on a silica substrate in which a microcrater is created by laser ablation. 
The sample and top mirror are the same as in Refs.\ \cite{Tomm2021,Tomm_direct2022}.
The much higher transmittance of the top mirror makes the cavity one-sided: the top mirror of the cavity is the main access port for incoming and outgoing light \cite{Tomm2021}. The semiconductor is mounted on a set of xyz-nanopositioners allowing full control over the cavity length ($z$), also the QD lateral position relative to the cavity centre ($xy$). A combination of a polarising beam-splitter and a half-wave plate gives full control over the polarisation of the input and output states. The cavity mode is frequency-split by $\delta_\text{cav}/(2\pi)=(\omega_\text{H}-\omega_\text{V})/(2\pi)=50$ \,GHz into two linearly- and orthogonally-polarised modes due to a small birefringence in the bottom mirror. We name these two polarisations $\textrm{H}$ and $\textrm{V}$, with detunings $\Delta\omega_\text{H/V}=\omega_\text{laser}- \omega_\text{H/V}$. The cavity modes have a loss rate $\kappa/(2\pi)$ = 28\,GHz.

We use a neutral exciton ($X^0$) in a QD. 
The QD was chosen according to two criteria: the exciton- and cavity-axes align reasonably well; the $X^0$ frequency lies in the intersection of the mirror stopbands.
The $X^0$ has a $V$-level energy structure: one ground state $\ket{g}$ with two excited states $\ket{a}$ and $\ket{b}$ [Fig.\,\ref{fig:fig1}\,(b)], with detunings $\Delta\omega_{a/b}= \omega_\text{laser} - \omega_{a/b}$. The two QD transitions are linearly polarised, mutually orthogonal, and split in frequency by $\delta_\text{QD}/(2\pi)= (\omega_a - \omega_b)/(2\pi)=2.3$\,GHz. The polarisation axes of the QD lie at an angle of $\theta=25.1^\circ$ relative to the axis of the cavity [Fig.\,\ref{fig:fig1}\,(b)]. Thus, the $\text{H}_\text{QD}$ transition $\ket{a}\leftrightarrow\ket{g}$  couples more to the H- than to the V-polarised cavity mode, and vice-versa for the $\text{V}_\text{QD}$ transition $\ket{b}\leftrightarrow\ket{g}$. When optimally coupled to the cavity, the QD transitions have a Purcell-enhanced decay rate $\Gamma/(2\pi)$ =1.65\,GHz (Purcell factor $F_{\rm P}$ = 11), giving a maximum coupling efficiency $\beta=F_{\rm P}/(F_{\rm P}+1)=92\%$.

We use two experimental configurations termed ``transmission" and ``reflection". We focus first on the transmission mode: Light with polarisation $\text{P} =  ( \text{H} + \text{V})/{\sqrt{2}}$ is input from Port 1 and interacts with the QD-cavity system. The output is collected in Port 2, with polarisation $\text{M} =  ( \text{H} - \text{V})/{\sqrt{2}}$ [Fig.~\ref{fig:fig1}(a)]. We thus measure light transmitted from the P to the M polarisation.

\begin{figure}[t!]
    \centering
    \includegraphics{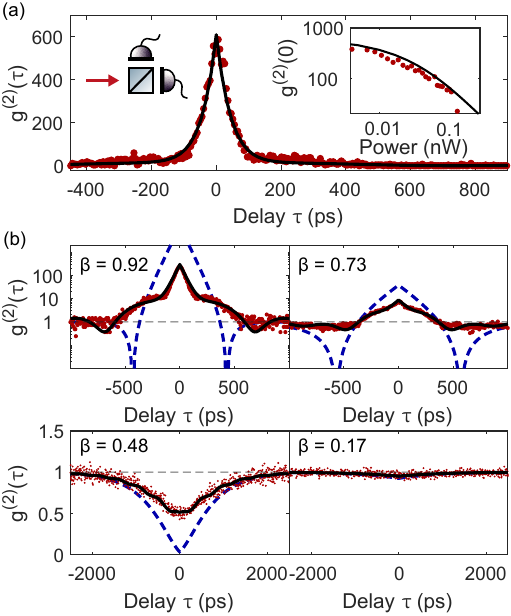}
    \caption{\textbf{Intensity correlations.} 
    \textbf{(a)} $g^{(2)}(\tau$) of the transmitted light for $\Delta\omega_\text{H}=0$ and $\Delta\omega_a=-0.14$\,GHz. The QD is positioned in the centre of the cavity ($\beta = 0.92$).
    The input power is 0.3 pW corresponding to 0.00067 photons per lifetime. The bin-size is 5 ps, integration time 1 hour; at large $\tau$, signal on average 0.24 counts-per-bin. The single-photon detector has efficiency 80\%, dark-count rate 100 Hz, and total timing jitter 50 ps.
    A bunching of 587 highlights the strong nonlinearity of the system. Inset shows $g^{(2)}(0)$ as a function of laser power, the black lines the theoretical model.
    \textbf{(b)} $g^{(2)}(\tau$) for different QD positions, i.e.\, different $\beta$-factors, for $\Delta\omega_\text{H}=0$ and $\Delta\omega_a/(2\pi)=-0.14$ GHz. 
    Input power 17 pW, bin-size and integration time as in (a); average counts-per-bin 8.2, 70, 414, 1700 for $\beta=0.92$, 0.73, 0.48 and 0.17, respectively.
    Bunching 
    turns into anti-bunching at $\beta \simeq 0.5$, and disappears for $\beta\rightarrow0$. The full model (solid black-line) is compared to the JC model (dashed blue-line).
    }
    \label{fig:fig2}
\end{figure}

\textit{Strong extinction.} 
The measured transmission as a function of QD detuning is shown in Fig.~\ref{fig:fig1}\,(d), illustrating an extinction of 99.2\%, an immediate demonstration of the efficient QD-cavity coupling.

Some features can be understood with a JC model, obtained by setting $\theta=0$ (ignoring the misalignment between the QD and cavity polarisations) and $|\delta_\text{cav}|/\kappa\to \infty$ (large cavity-mode separation). In this case, only H-polarised light interacts with the QD-cavity; V-polarised light is perfectly reflected. This model represents the canonical system, a two-level emitter coupled to a single cavity-mode~\cite{rice_single-atom_1988, auffeves-garnier_giant_2007}. 

We consider first the case when the QD is far out of resonance [Fig.~\ref{fig:fig1}\,(c), far left/right]. When $\Delta \omega_{\text H}$ is swept across zero, the phase of the reflected H-polarised light winds around $2\pi$ while the V-polarised light remains unchanged. On resonance ($\Delta \omega_\text{H} = 0 $), the H-polarised light obtains a phase shift of $\pi$, turning P-polarised light into M-polarised light and resulting in a transmission close to unity. This explains the peaks in Fig.~\ref{fig:fig1}\,(c) when the laser is on resonance with one of the cavity modes (while the QD is out of resonance). Next, we consider sweeping the detuning of the QD $\Delta \omega_a$ at $\Delta \omega_\text{H}=0$ [Fig.~\ref{fig:fig1}\,(d)]. Crossing $\Delta \omega_a=0$, the phase of the reflected H-polarised light again winds around $2\pi$. This results in a dip in transmission; in the absence of dissipation, we expect perfect extinction ($T=0$) on resonance ($\Delta \omega_a=\Delta \omega_\text{H}=0$). Dissipation reduces the transmission to $T=(1-\beta)^2$ \cite{suppmat, auffeves-garnier_giant_2007}. 

 Features of the transmission that are not captured by the JC model are the double-dip structure [Fig.~\ref{fig:fig1}\,(d)], and the shift of the maximal extinction to $\Delta\omega_a/(2\pi)\approx - 0.31 \text{GHz}$. The double-dip arises because in the presence of two QD transitions, sweeping $\Delta\omega_a$ results in the crossing of two resonances. The phase of the reflected H-polarised light then winds around $4\pi$, resulting in two dips. Considering the full three-level system, the theory shows excellent agreement with experiment [Fig~\ref{fig:fig1}(d)].
 
The maximum extinction is strongly dependent on the input power. With increasing laser power, the transmission dip disappears [inset, Fig.~\ref{fig:fig1}(d)]. This nonlinear response is a consequence of the saturation of the quantum emitter. We extract a saturation power of $P_{\textrm{sat}}$~=~1.8\,nW~\footnote{The theory predicts $P_{\textrm{sat}}=1.5$ nW; the mismatch is attributed to losses \cite{suppmat}.}, corresponding to a flux of 0.4 photons per QD lifetime. 
 
\textit{Giant and tunable nonlinearity.} We now demonstrate the ability of this cavity-QED setup to manipulate the statistics of the transmitted light.
To this end, we consider the second-order correlation function $g^{(2)}(\tau)$, with $\tau$ the delay between detection events. We observe very strong bunching, $g^{(2)}(\tau=0)$ = 587  [Fig.\,\ref{fig:fig2}\,(a)] for very low input-power and an optimally-coupled QD. To our knowledge, this is the largest photon-bunching due to a nonlinearity observed to date. Such a high bunching demands low dephasing, $\beta\simeq 1$, and high signal-to-laser-background ratio. Tuning $\beta$ results in a change from strong bunching to anti-bunching [Fig. \ref{fig:fig2} (b)], demonstrating wide control over the statistics of the transmitted light. In the experiment, $\beta$ is tuned by laterally moving the QD relative to the cavity centre.

The $g^{(2)}(\tau)$ function exhibiting giant bunching can be explained by the evolution of the M-polarised cavity field. In the bad-cavity, weak-drive limit, the cavity field is described by a pure quantum state at all times~\cite{Carmichael2008, Carmichael1989PRA, rice_single-atom_1988,suppmat} 
\begin{equation}
    \label{eq:carmichael}
    \ket{\psi}_\tau = \ket{\alpha} -i\sqrt{\frac{\Gamma}{2\kappa}}\langle \hat{\sigma}_\text{M}\rangle_\tau\hat{a}_\text{M}^\dagger\ket{\alpha}.
\end{equation}
Here, $\ket{\alpha}$ denotes the coherent state describing the M-mode in the absence of the QD. The second term describes the effect of the QD; $\hat{a}_\text{M}^\dagger$ denotes the creation operator of the M-mode, and $ \hat{\sigma}_\text{M}$ the QD transition coupled to the M polarisation. Equation~\eqref{eq:carmichael} provides the correct averages for any normal-ordered observable to leading order in the external drive. We refer to this state as \textit{Rice-Carmichael (RC) state}~\cite{rice_single-atom_1988}. The $g^{(2)}$-function may then be understood as follows. Detecting a photon alters the field in the cavity, which then regresses to its steady state. The time-dependent average field gives $g^{(2)}(\tau) = |\langle{\hat{a}_\text{M}}\rangle_\tau/\langle \hat{a}_\text{M} \rangle_\infty|^2$, where the average is relative to \ $\ket{\psi}_\tau$~\cite{suppmat, rice_single-atom_1988}. Thus, $g^{(2)}(\tau)$ larger (smaller) than one is observed whenever the average field is stronger (weaker) than in the steady state.

To explain the key aspects, we consider the RC state for the JC model. In this case, we find 
\begin{equation}
\label{eq:avgat}
    \frac{\langle{\hat{a}_\text{M}}\rangle_\tau}{\langle \hat{a}_\text{M} \rangle_\infty} = 1-\frac{\beta^2}{(1-\beta)^2}e^{-\frac{\gamma \tau}{2(1-\beta)}},
\end{equation}
where $\gamma = 2\Gamma/F_{\rm P}$ denotes the dissipation rate of the QD. This shows that the photon statistics can be modified by tuning $\beta$.
At $\beta\simeq0$ the cavity field remains close to a coherent state while for $\beta\simeq1$, the contribution to Eq.~\eqref{eq:avgat} stemming from the QD yields an amplified number of photons.

The transition from bunching to anti-bunching [Fig.\,\ref{fig:fig2}\,(b)] may qualitatively be understood by considering the steady state in the Fock basis $\braket{n}{\psi}_\infty=\alpha^n(1-\beta n)/\sqrt{n!}$. 
At $\beta=1$ the single-photon component in the cavity vanishes and is thus perfectly reflected [sketch Fig.\,\ref{fig:fig1}\,(a)]. Crucially, multi-photon components are present in the cavity, leading to giant bunching in the transmitted light. In particular, $|\braket{2}{\psi}_\infty|$ remains unchanged. In contrast, for $\beta=1/2$, the two-photon component in the cavity vanishes, yielding perfect anti-bunching $g^{(2)}(0)=0$. Similarly, tuning $\beta=1/n$ allows the $n$-photon component to be suppressed. 

Equation~\eqref{eq:avgat} implies that the average field in the cavity changes sign upon photodetection if $\beta>1/2$. When the field subsequently crosses zero, $g^{(2)}$ vanishes [Fig.~\ref{fig:fig2}\,(b)]. 

Features not captured by the JC model include the shoulders in Fig.~\ref{fig:fig2}\,(b), which are related to the cavity and QD splittings, $\delta_\text{QD/cav}\neq0$. Furthermore, the anti-bunching is limited by the cavity-mode splitting $\delta_\text{cav}\neq0$ and the polarisations' misalignment, $\theta\neq0$, such that $g^{(2)}(0)\simeq 0.5$ at $\beta\simeq 0.5$. Our full model shows excellent agreement with these features.

\begin{figure}[t!]
    \centering
    \includegraphics{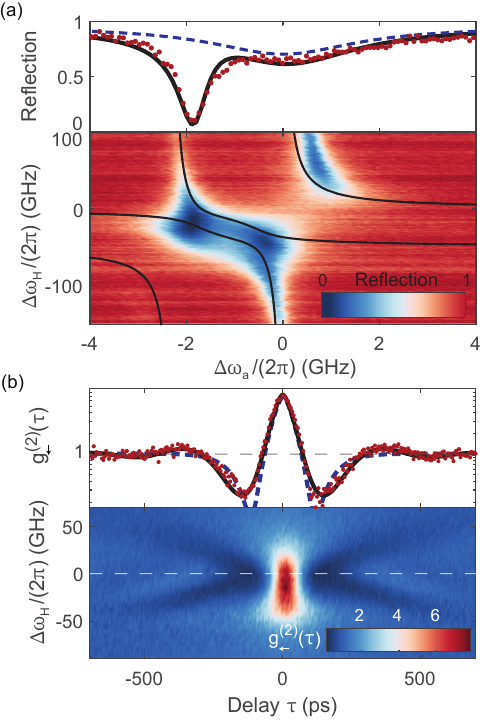}
    \caption{\textbf{Reflection mode} 
    \textbf{(a)} 
    Top: Reflection as a function of QD detuning for $\Delta\omega_\text{H}=0$ and a power well below saturation. The weakly coupled QD transition results in a much deeper dip than the strongly coupled transition.
    Bottom: Reflection as a function of QD and cavity detuning. The black lines show the resonances between the laser frequency and the lowest transition frequencies of the Hamiltonian, see \cite{suppmat}.
    \textbf{(b)}
    Top: $g^{(2)}(\tau$) at $\Delta\omega_\text{H} =0, \Delta\omega_a/(2\pi)=-0.14$ GHz. Correlation followed by anti-correlation is observed.
    Bottom: $g^{(2)}(\tau$) versus $\Delta\omega_{\rm H}$. The dashed line marks the cut-through, $\Delta \omega_{\rm H} = 0$. 
    In the top panels, the full theoretical model (solid black-line) is compared to the JC model (dashed blue-line).}
    \label{fig:fig3}
\end{figure}

\textit{Photons are correlated in time.}
We now turn to the reflection mode experiment~\cite{lodahl_chiral_2017, antoniadis_chiral_2022}. In this case, the cavity is driven by H-polarised light and the reflected light in the same polarisation is measured. A 99:1 beam-splitter separates the input light from the output. Figure\,\ref{fig:fig3}\,(a) shows the reflected signal $R_\leftarrow$. Due to the coupling to the QD, photons either dissipate or change polarisation, reducing the reflection. Interestingly, the less-coupled QD transition results in a stronger reduction of the reflection [upper panel, Fig.\,\ref{fig:fig3}\,(a)]. This is explained by the JC model, obtained for $\theta=0$ in the reflection mode. Specifically, for $\Delta\omega_{\rm H}=\Delta\omega_a=0$, $R_\leftarrow=(1-2\beta)^2$; then, for $\beta= 0.92$, we have a small dip in the reflection [blue line, Fig.~\ref{fig:fig3}\,(a)]. The less-coupled transition results in a pronounced dip: the reduced coupling is equivalent to a smaller $\beta$.

Turning to $g^{(2)}_\leftarrow (\tau)$, Fig.~\ref{fig:fig3}\,(b), we find a peak at zero delay $g^{(2)}_\leftarrow(0)=7.3$, followed by a dip $g^{(2)}_\leftarrow(133\text{ps})=  0.25$. Photons are thus correlated at short delays and anti-correlated at longer delays. Thus, it is much more likely to observe them close to each other. This is consistent with the formation of bound states: photons are pulled together in time, forming a highly correlated state \cite{Tomm_direct2022,Mahmoodian2020}. A similar effect is also predicted in the transmission mode [Fig.~\ref{fig:fig2}\,(b)], but there it lies below the noise floor due to the small transmission. The intensity correlations in reflection mode can be described using an RC state: $g^{(2)}_\leftarrow(\tau) = |\langle{\hat{b}_\text{H}}\rangle_\tau/\langle \hat{b}_\text{H} \rangle_\infty|^2$, where the averages are computed from the RC state, and $\hat{b}_\text{H}$ describes the cavity field displaced by the light that is directly reflected. The JC model again allows the qualitative features of the data to be understood. For $\Delta\omega_{\rm H}=\Delta\omega_a=0$, 
\begin{equation}
\label{eq:avgat2}
    \frac{\langle{\hat{b}_\text{H}}\rangle_\tau}{\langle \hat{b}_\text{H} \rangle_\infty} = 1-\frac{(2\beta)^2}{(1-2\beta)^2}e^{-\frac{\gamma \tau}{2(1-\beta)}}.
\end{equation}
The field changes sign upon detecting a photon if $\beta>1/4$. When regressing to the steady state, we must have $\langle{\hat{b}_\text{H}}\rangle=0$, yielding a vanishing $g^{(2)}_{\leftarrow}$ and explaining the anti-correlation at finite delays \cite{suppmat}.

\textit{Conclusion and outlook.} We have established efficient and coherent coupling between a QD and a microcavity. As a result, the QD behaves radically differently depending on the number of photons. This leads to a huge bunching, as only multi-photon states are transmitted. We find excellent agreement between experiment and theory by going beyond the single-mode JC model and including two QD transitions and two cavity modes in our model. The connection between the RC-state and the $g^{(2)}$ suggests an intuitive understanding of photon statistics through the field's phase-space.

The observed photon-number discriminating interaction enables photon-photon interactions at the single-photon limit and may find application in creating photonic bound states \cite{Stiesdal2018, Liang2018, Tomm_direct2022} and exotic photonic states \cite{Prasad2020}, in establishing direction-dependent phase-shifts \cite{lodahl_chiral_2017}, and in studying many-body phenomena \cite{noh_quantum_2016} in a controllable setting. The overall transmission of the setup, from the input fibre to the output fibre, is high, about 50\% \cite{Tomm2021}. 
A promising avenue is to mimic the interaction between photons and multiple quantum emitters in series e.g.\ via time-delayed feedback. This allows the generation of exotic bound states involving many photons \cite{Mahmoodian2020}.
Furthermore, this system may find application as a photon sorter \cite{ralph_photon_2015, yang_deterministic_2022, Witthaut2012, shomroni_all-optical_2014} or a photon-photon quantum gate \cite{chen_two-photon_2021, brod_passive_2016}.

\begin{acknowledgments}
\textit{Acknowledgements.} We thank Mark R.\ Hogg, Gabriel T.\ Landi and Aaron Daniel for stimulating discussions. We acknowledge financial support from Horizon-2020 FET-Open Project QLUSTER, Swiss National Science Foundation project 200020\_204069, Eccellenza Professorial Fellowship PCEFP2\_194268, and NCCR QSIT. A.J. acknowledges support from the European Unions Horizon 2020 Research and Innovation Programme under Marie Sk\l{}odowska-Curie grant agreement no. 840453 (HiFig). S.R.V., R.S., A.L. and A.D.W. acknowledge support from DFH/UFA CDFA05-06, DFG TRR160, DFG project 383065199 and BMBF-QR.X Project 16KISQ009.
\end{acknowledgments}

\bibliography{references}

\begin{thebibliography}{50}%
\makeatletter
\providecommand \@ifxundefined [1]{%
 \@ifx{#1\undefined}
}%
\providecommand \@ifnum [1]{%
 \ifnum #1\expandafter \@firstoftwo
 \else \expandafter \@secondoftwo
 \fi
}%
\providecommand \@ifx [1]{%
 \ifx #1\expandafter \@firstoftwo
 \else \expandafter \@secondoftwo
 \fi
}%
\providecommand \natexlab [1]{#1}%
\providecommand \enquote  [1]{``#1''}%
\providecommand \bibnamefont  [1]{#1}%
\providecommand \bibfnamefont [1]{#1}%
\providecommand \citenamefont [1]{#1}%
\providecommand \href@noop [0]{\@secondoftwo}%
\providecommand \href [0]{\begingroup \@sanitize@url \@href}%
\providecommand \@href[1]{\@@startlink{#1}\@@href}%
\providecommand \@@href[1]{\endgroup#1\@@endlink}%
\providecommand \@sanitize@url [0]{\catcode `\\12\catcode `\$12\catcode
  `\&12\catcode `\#12\catcode `\^12\catcode `\_12\catcode `\%12\relax}%
\providecommand \@@startlink[1]{}%
\providecommand \@@endlink[0]{}%
\providecommand \url  [0]{\begingroup\@sanitize@url \@url }%
\providecommand \@url [1]{\endgroup\@href {#1}{\urlprefix }}%
\providecommand \urlprefix  [0]{URL }%
\providecommand \Eprint [0]{\href }%
\providecommand \doibase [0]{https://doi.org/}%
\providecommand \selectlanguage [0]{\@gobble}%
\providecommand \bibinfo  [0]{\@secondoftwo}%
\providecommand \bibfield  [0]{\@secondoftwo}%
\providecommand \translation [1]{[#1]}%
\providecommand \BibitemOpen [0]{}%
\providecommand \bibitemStop [0]{}%
\providecommand \bibitemNoStop [0]{.\EOS\space}%
\providecommand \EOS [0]{\spacefactor3000\relax}%
\providecommand \BibitemShut  [1]{\csname bibitem#1\endcsname}%
\let\auto@bib@innerbib\@empty
\bibitem [{\citenamefont {Chang}\ \emph {et~al.}(2014)\citenamefont {Chang},
  \citenamefont {Vuletić},\ and\ \citenamefont {Lukin}}]{chang_quantum_2014}%
  \BibitemOpen
  \bibfield  {author} {\bibinfo {author} {\bibfnamefont {D.~E.}\ \bibnamefont
  {Chang}}, \bibinfo {author} {\bibfnamefont {V.}~\bibnamefont {Vuletić}},\
  and\ \bibinfo {author} {\bibfnamefont {M.~D.}\ \bibnamefont {Lukin}},\
  }\bibfield  {title} {\bibinfo {title} {Quantum nonlinear optics — photon by
  photon},\ }\href {https://doi.org/10.1038/nphoton.2014.192} {\bibfield
  {journal} {\bibinfo  {journal} {Nat. Photon.}\ }\textbf {\bibinfo {volume}
  {8}},\ \bibinfo {pages} {685} (\bibinfo {year} {2014})}\BibitemShut {NoStop}%
\bibitem [{\citenamefont {Hartmann}(2016)}]{hartmann_quantum_2016}%
  \BibitemOpen
  \bibfield  {author} {\bibinfo {author} {\bibfnamefont {M.~J.}\ \bibnamefont
  {Hartmann}},\ }\bibfield  {title} {{\selectlanguage {english}\bibinfo {title}
  {Quantum simulation with interacting photons}},\ }\href
  {https://doi.org/10.1088/2040-8978/18/10/104005} {\bibfield  {journal}
  {\bibinfo  {journal} {J. Opt.}\ }\textbf {\bibinfo {volume} {18}},\ \bibinfo
  {pages} {104005} (\bibinfo {year} {2016})}\BibitemShut {NoStop}%
\bibitem [{\citenamefont {Chang}\ \emph {et~al.}(2018)\citenamefont {Chang},
  \citenamefont {Douglas}, \citenamefont {Gonz\'alez-Tudela}, \citenamefont
  {Hung},\ and\ \citenamefont {Kimble}}]{Chang2018RMP}%
  \BibitemOpen
  \bibfield  {author} {\bibinfo {author} {\bibfnamefont {D.~E.}\ \bibnamefont
  {Chang}}, \bibinfo {author} {\bibfnamefont {J.~S.}\ \bibnamefont {Douglas}},
  \bibinfo {author} {\bibfnamefont {A.}~\bibnamefont {Gonz\'alez-Tudela}},
  \bibinfo {author} {\bibfnamefont {C.-L.}\ \bibnamefont {Hung}},\ and\
  \bibinfo {author} {\bibfnamefont {H.~J.}\ \bibnamefont {Kimble}},\ }\bibfield
   {title} {\bibinfo {title} {Colloquium: Quantum matter built from nanoscopic
  lattices of atoms and photons},\ }\href
  {https://doi.org/10.1103/RevModPhys.90.031002} {\bibfield  {journal}
  {\bibinfo  {journal} {Rev. Mod. Phys.}\ }\textbf {\bibinfo {volume} {90}},\
  \bibinfo {pages} {031002} (\bibinfo {year} {2018})}\BibitemShut {NoStop}%
\bibitem [{\citenamefont {Shen}\ and\ \citenamefont
  {Fan}(2007{\natexlab{a}})}]{shen_strongly_2007}%
  \BibitemOpen
  \bibfield  {author} {\bibinfo {author} {\bibfnamefont {J.-T.}\ \bibnamefont
  {Shen}}\ and\ \bibinfo {author} {\bibfnamefont {S.}~\bibnamefont {Fan}},\
  }\bibfield  {title} {\bibinfo {title} {Strongly correlated multiparticle
  transport in one dimension through a quantum impurity},\ }\href
  {https://doi.org/10.1103/PhysRevA.76.062709} {\bibfield  {journal} {\bibinfo
  {journal} {Phys. Rev. A}\ }\textbf {\bibinfo {volume} {76}},\ \bibinfo
  {pages} {062709} (\bibinfo {year} {2007}{\natexlab{a}})}\BibitemShut
  {NoStop}%
\bibitem [{\citenamefont {Sørensen}\ \emph {et~al.}(2016)\citenamefont
  {Sørensen}, \citenamefont {Béguin}, \citenamefont {Kluge}, \citenamefont
  {Iakoupov}, \citenamefont {Sørensen}, \citenamefont {Müller}, \citenamefont
  {Polzik},\ and\ \citenamefont {Appel}}]{sorensen_coherent_2016}%
  \BibitemOpen
  \bibfield  {author} {\bibinfo {author} {\bibfnamefont {H.}~\bibnamefont
  {Sørensen}}, \bibinfo {author} {\bibfnamefont {J.-B.}\ \bibnamefont
  {Béguin}}, \bibinfo {author} {\bibfnamefont {K.}~\bibnamefont {Kluge}},
  \bibinfo {author} {\bibfnamefont {I.}~\bibnamefont {Iakoupov}}, \bibinfo
  {author} {\bibfnamefont {A.}~\bibnamefont {Sørensen}}, \bibinfo {author}
  {\bibfnamefont {J.}~\bibnamefont {Müller}}, \bibinfo {author} {\bibfnamefont
  {E.}~\bibnamefont {Polzik}},\ and\ \bibinfo {author} {\bibfnamefont
  {J.}~\bibnamefont {Appel}},\ }\bibfield  {title} {\bibinfo {title} {Coherent
  {Backscattering} of {Light} {Off} {One}-{Dimensional} {Atomic} {Strings}},\
  }\href {https://doi.org/10.1103/PhysRevLett.117.133604} {\bibfield  {journal}
  {\bibinfo  {journal} {Phys. Rev. Lett.}\ }\textbf {\bibinfo {volume} {117}},\
  \bibinfo {pages} {133604} (\bibinfo {year} {2016})}\BibitemShut {NoStop}%
\bibitem [{\citenamefont {Prasad}\ \emph {et~al.}(2020)\citenamefont {Prasad},
  \citenamefont {Hinney}, \citenamefont {Mahmoodian}, \citenamefont {Hammerer},
  \citenamefont {Rind}, \citenamefont {Schneeweiss}, \citenamefont
  {S{\o}rensen}, \citenamefont {Volz},\ and\ \citenamefont
  {Rauschenbeutel}}]{Prasad2020}%
  \BibitemOpen
  \bibfield  {author} {\bibinfo {author} {\bibfnamefont {A.~S.}\ \bibnamefont
  {Prasad}}, \bibinfo {author} {\bibfnamefont {J.}~\bibnamefont {Hinney}},
  \bibinfo {author} {\bibfnamefont {S.}~\bibnamefont {Mahmoodian}}, \bibinfo
  {author} {\bibfnamefont {K.}~\bibnamefont {Hammerer}}, \bibinfo {author}
  {\bibfnamefont {S.}~\bibnamefont {Rind}}, \bibinfo {author} {\bibfnamefont
  {P.}~\bibnamefont {Schneeweiss}}, \bibinfo {author} {\bibfnamefont {A.~S.}\
  \bibnamefont {S{\o}rensen}}, \bibinfo {author} {\bibfnamefont
  {J.}~\bibnamefont {Volz}},\ and\ \bibinfo {author} {\bibfnamefont
  {A.}~\bibnamefont {Rauschenbeutel}},\ }\bibfield  {title} {\bibinfo {title}
  {Correlating photons using the collective nonlinear response of atoms weakly
  coupled to an optical mode},\ }\href
  {https://doi.org/10.1038/s41566-020-0692-z} {\bibfield  {journal} {\bibinfo
  {journal} {Nat. Photon.}\ }\textbf {\bibinfo {volume} {14}},\ \bibinfo
  {pages} {719} (\bibinfo {year} {2020})}\BibitemShut {NoStop}%
\bibitem [{\citenamefont {Arcari}\ \emph {et~al.}(2014)\citenamefont {Arcari},
  \citenamefont {Söllner}, \citenamefont {Javadi}, \citenamefont
  {Lindskov~Hansen}, \citenamefont {Mahmoodian}, \citenamefont {Liu},
  \citenamefont {Thyrrestrup}, \citenamefont {Lee}, \citenamefont {Song},
  \citenamefont {Stobbe},\ and\ \citenamefont
  {Lodahl}}]{arcari_near-unity_2014}%
  \BibitemOpen
  \bibfield  {author} {\bibinfo {author} {\bibfnamefont {M.}~\bibnamefont
  {Arcari}}, \bibinfo {author} {\bibfnamefont {I.}~\bibnamefont {Söllner}},
  \bibinfo {author} {\bibfnamefont {A.}~\bibnamefont {Javadi}}, \bibinfo
  {author} {\bibfnamefont {S.}~\bibnamefont {Lindskov~Hansen}}, \bibinfo
  {author} {\bibfnamefont {S.}~\bibnamefont {Mahmoodian}}, \bibinfo {author}
  {\bibfnamefont {J.}~\bibnamefont {Liu}}, \bibinfo {author} {\bibfnamefont
  {H.}~\bibnamefont {Thyrrestrup}}, \bibinfo {author} {\bibfnamefont
  {E.}~\bibnamefont {Lee}}, \bibinfo {author} {\bibfnamefont {J.}~\bibnamefont
  {Song}}, \bibinfo {author} {\bibfnamefont {S.}~\bibnamefont {Stobbe}},\ and\
  \bibinfo {author} {\bibfnamefont {P.}~\bibnamefont {Lodahl}},\ }\bibfield
  {title} {\bibinfo {title} {Near-{Unity} {Coupling} {Efficiency} of a
  {Quantum} {Emitter} to a {Photonic} {Crystal} {Waveguide}},\ }\href
  {https://doi.org/10.1103/PhysRevLett.113.093603} {\bibfield  {journal}
  {\bibinfo  {journal} {Phys. Rev. Lett.}\ }\textbf {\bibinfo {volume} {113}},\
  \bibinfo {pages} {093603} (\bibinfo {year} {2014})}\BibitemShut {NoStop}%
\bibitem [{\citenamefont {Hallett}\ \emph {et~al.}(2018)\citenamefont
  {Hallett}, \citenamefont {Foster}, \citenamefont {Hurst}, \citenamefont
  {Royall}, \citenamefont {Kok}, \citenamefont {Clarke}, \citenamefont
  {Itskevich}, \citenamefont {Fox}, \citenamefont {Skolnick},\ and\
  \citenamefont {Wilson}}]{Hallett2018}%
  \BibitemOpen
  \bibfield  {author} {\bibinfo {author} {\bibfnamefont {D.}~\bibnamefont
  {Hallett}}, \bibinfo {author} {\bibfnamefont {A.~P.}\ \bibnamefont {Foster}},
  \bibinfo {author} {\bibfnamefont {D.~L.}\ \bibnamefont {Hurst}}, \bibinfo
  {author} {\bibfnamefont {B.}~\bibnamefont {Royall}}, \bibinfo {author}
  {\bibfnamefont {P.}~\bibnamefont {Kok}}, \bibinfo {author} {\bibfnamefont
  {E.}~\bibnamefont {Clarke}}, \bibinfo {author} {\bibfnamefont {I.~E.}\
  \bibnamefont {Itskevich}}, \bibinfo {author} {\bibfnamefont {A.~M.}\
  \bibnamefont {Fox}}, \bibinfo {author} {\bibfnamefont {M.~S.}\ \bibnamefont
  {Skolnick}},\ and\ \bibinfo {author} {\bibfnamefont {L.~R.}\ \bibnamefont
  {Wilson}},\ }\bibfield  {title} {\bibinfo {title} {Electrical control of
  nonlinear quantum optics in a nano-photonic waveguide},\ }\href
  {https://doi.org/10.1364/OPTICA.5.000644} {\bibfield  {journal} {\bibinfo
  {journal} {Optica}\ }\textbf {\bibinfo {volume} {5}},\ \bibinfo {pages} {644}
  (\bibinfo {year} {2018})}\BibitemShut {NoStop}%
\bibitem [{\citenamefont {Le~Jeannic}\ \emph {et~al.}(2021)\citenamefont
  {Le~Jeannic}, \citenamefont {Ramos}, \citenamefont {Simonsen}, \citenamefont
  {Pregnolato}, \citenamefont {Liu}, \citenamefont {Schott}, \citenamefont
  {Wieck}, \citenamefont {Ludwig}, \citenamefont {Rotenberg}, \citenamefont
  {Garc\'ia-Ripoll},\ and\ \citenamefont {Lodahl}}]{LeJeannic2021PRL}%
  \BibitemOpen
  \bibfield  {author} {\bibinfo {author} {\bibfnamefont {H.}~\bibnamefont
  {Le~Jeannic}}, \bibinfo {author} {\bibfnamefont {T.}~\bibnamefont {Ramos}},
  \bibinfo {author} {\bibfnamefont {S.~F.}\ \bibnamefont {Simonsen}}, \bibinfo
  {author} {\bibfnamefont {T.}~\bibnamefont {Pregnolato}}, \bibinfo {author}
  {\bibfnamefont {Z.}~\bibnamefont {Liu}}, \bibinfo {author} {\bibfnamefont
  {R.}~\bibnamefont {Schott}}, \bibinfo {author} {\bibfnamefont {A.~D.}\
  \bibnamefont {Wieck}}, \bibinfo {author} {\bibfnamefont {A.}~\bibnamefont
  {Ludwig}}, \bibinfo {author} {\bibfnamefont {N.}~\bibnamefont {Rotenberg}},
  \bibinfo {author} {\bibfnamefont {J.~J.}\ \bibnamefont {Garc\'ia-Ripoll}},\
  and\ \bibinfo {author} {\bibfnamefont {P.}~\bibnamefont {Lodahl}},\
  }\bibfield  {title} {\bibinfo {title} {Experimental reconstruction of the
  few-photon nonlinear scattering matrix from a single quantum dot in a
  nanophotonic waveguide},\ }\href
  {https://journals.aps.org/prl/abstract/10.1103/PhysRevLett.126.023603}
  {\bibfield  {journal} {\bibinfo  {journal} {Phys. Rev. Lett.}\ }\textbf
  {\bibinfo {volume} {126}},\ \bibinfo {pages} {023603} (\bibinfo {year}
  {2021})}\BibitemShut {NoStop}%
\bibitem [{\citenamefont {Le~Jeannic}\ \emph {et~al.}(2022)\citenamefont
  {Le~Jeannic}, \citenamefont {Tiranov}, \citenamefont {Carolan}, \citenamefont
  {Ramos}, \citenamefont {Wang}, \citenamefont {Appel}, \citenamefont {Scholz},
  \citenamefont {Wieck}, \citenamefont {Ludwig}, \citenamefont {Rotenberg},
  \citenamefont {Midolo}, \citenamefont {Garc\'ia-Ripoll}, \citenamefont
  {S{\o}rensen},\ and\ \citenamefont {Lodahl}}]{LeJeannic2021}%
  \BibitemOpen
  \bibfield  {author} {\bibinfo {author} {\bibfnamefont {H.}~\bibnamefont
  {Le~Jeannic}}, \bibinfo {author} {\bibfnamefont {A.}~\bibnamefont {Tiranov}},
  \bibinfo {author} {\bibfnamefont {J.}~\bibnamefont {Carolan}}, \bibinfo
  {author} {\bibfnamefont {T.}~\bibnamefont {Ramos}}, \bibinfo {author}
  {\bibfnamefont {Y.}~\bibnamefont {Wang}}, \bibinfo {author} {\bibfnamefont
  {M.~H.}\ \bibnamefont {Appel}}, \bibinfo {author} {\bibfnamefont
  {S.}~\bibnamefont {Scholz}}, \bibinfo {author} {\bibfnamefont {A.~D.}\
  \bibnamefont {Wieck}}, \bibinfo {author} {\bibfnamefont {A.}~\bibnamefont
  {Ludwig}}, \bibinfo {author} {\bibfnamefont {N.}~\bibnamefont {Rotenberg}},
  \bibinfo {author} {\bibfnamefont {L.}~\bibnamefont {Midolo}}, \bibinfo
  {author} {\bibfnamefont {J.~J.}\ \bibnamefont {Garc\'ia-Ripoll}}, \bibinfo
  {author} {\bibfnamefont {A.~S.}\ \bibnamefont {S{\o}rensen}},\ and\ \bibinfo
  {author} {\bibfnamefont {P.}~\bibnamefont {Lodahl}},\ }\bibfield  {title}
  {\bibinfo {title} {Dynamical photon-photon interaction mediated by a quantum
  emitter},\ }\href {https://www.nature.com/articles/s41567-022-01720-x}
  {\bibfield  {journal} {\bibinfo  {journal} {Nat. Phys.}\ }\textbf {\bibinfo
  {volume} {18}},\ \bibinfo {pages} {1191–1195} (\bibinfo {year}
  {2022})}\BibitemShut {NoStop}%
\bibitem [{\citenamefont {Goban}\ \emph {et~al.}(2014)\citenamefont {Goban},
  \citenamefont {Hung}, \citenamefont {Yu}, \citenamefont {Hood}, \citenamefont
  {Muniz}, \citenamefont {Lee}, \citenamefont {Martin}, \citenamefont
  {McClung}, \citenamefont {Choi}, \citenamefont {Chang}, \citenamefont
  {Painter},\ and\ \citenamefont {Kimble}}]{goban_atomlight_2014}%
  \BibitemOpen
  \bibfield  {author} {\bibinfo {author} {\bibfnamefont {A.}~\bibnamefont
  {Goban}}, \bibinfo {author} {\bibfnamefont {C.-L.}\ \bibnamefont {Hung}},
  \bibinfo {author} {\bibfnamefont {S.-P.}\ \bibnamefont {Yu}}, \bibinfo
  {author} {\bibfnamefont {J.~D.}\ \bibnamefont {Hood}}, \bibinfo {author}
  {\bibfnamefont {J.~A.}\ \bibnamefont {Muniz}}, \bibinfo {author}
  {\bibfnamefont {J.~H.}\ \bibnamefont {Lee}}, \bibinfo {author} {\bibfnamefont
  {M.~J.}\ \bibnamefont {Martin}}, \bibinfo {author} {\bibfnamefont {A.~C.}\
  \bibnamefont {McClung}}, \bibinfo {author} {\bibfnamefont {K.~S.}\
  \bibnamefont {Choi}}, \bibinfo {author} {\bibfnamefont {D.~E.}\ \bibnamefont
  {Chang}}, \bibinfo {author} {\bibfnamefont {O.}~\bibnamefont {Painter}},\
  and\ \bibinfo {author} {\bibfnamefont {H.~J.}\ \bibnamefont {Kimble}},\
  }\bibfield  {title} {{\selectlanguage {english}\bibinfo {title} {Atom–light
  interactions in photonic crystals}},\ }\href
  {https://doi.org/10.1038/ncomms4808} {\bibfield  {journal} {\bibinfo
  {journal} {Nat. Commun.}\ }\textbf {\bibinfo {volume} {5}},\ \bibinfo {pages}
  {3808} (\bibinfo {year} {2014})}\BibitemShut {NoStop}%
\bibitem [{\citenamefont {Tiecke}\ \emph {et~al.}(2014)\citenamefont {Tiecke},
  \citenamefont {Thompson}, \citenamefont {de~Leon}, \citenamefont {Liu},
  \citenamefont {Vuleti{\'c}},\ and\ \citenamefont {Lukin}}]{Tiecke2014Nature}%
  \BibitemOpen
  \bibfield  {author} {\bibinfo {author} {\bibfnamefont {T.~G.}\ \bibnamefont
  {Tiecke}}, \bibinfo {author} {\bibfnamefont {J.~D.}\ \bibnamefont
  {Thompson}}, \bibinfo {author} {\bibfnamefont {N.~P.}\ \bibnamefont
  {de~Leon}}, \bibinfo {author} {\bibfnamefont {L.~R.}\ \bibnamefont {Liu}},
  \bibinfo {author} {\bibfnamefont {V.}~\bibnamefont {Vuleti{\'c}}},\ and\
  \bibinfo {author} {\bibfnamefont {M.}~\bibnamefont {Lukin}},\ }\bibfield
  {title} {\bibinfo {title} {Nanophotonic quantum phase switch with a single
  atom},\ }\href {https://doi.org/10.1038/nature13188} {\bibfield  {journal}
  {\bibinfo  {journal} {Nature}\ }\textbf {\bibinfo {volume} {508}},\ \bibinfo
  {pages} {241} (\bibinfo {year} {2014})}\BibitemShut {NoStop}%
\bibitem [{\citenamefont {Luan}\ \emph {et~al.}(2020)\citenamefont {Luan},
  \citenamefont {Béguin}, \citenamefont {Burgers}, \citenamefont {Qin},
  \citenamefont {Yu},\ and\ \citenamefont {Kimble}}]{luan_integration_2020}%
  \BibitemOpen
  \bibfield  {author} {\bibinfo {author} {\bibfnamefont {X.}~\bibnamefont
  {Luan}}, \bibinfo {author} {\bibfnamefont {J.-B.}\ \bibnamefont {Béguin}},
  \bibinfo {author} {\bibfnamefont {A.~P.}\ \bibnamefont {Burgers}}, \bibinfo
  {author} {\bibfnamefont {Z.}~\bibnamefont {Qin}}, \bibinfo {author}
  {\bibfnamefont {S.-P.}\ \bibnamefont {Yu}},\ and\ \bibinfo {author}
  {\bibfnamefont {H.~J.}\ \bibnamefont {Kimble}},\ }\bibfield  {title}
  {{\selectlanguage {english}\bibinfo {title} {The {Integration} of {Photonic}
  {Crystal} {Waveguides} with {Atom} {Arrays} in {Optical} {Tweezers}}},\
  }\href {https://doi.org/10.1002/qute.202000008} {\bibfield  {journal}
  {\bibinfo  {journal} {Adv. Quantum Technol.}\ }\textbf {\bibinfo {volume}
  {3}},\ \bibinfo {pages} {2000008} (\bibinfo {year} {2020})}\BibitemShut
  {NoStop}%
\bibitem [{\citenamefont {Takahashi}\ \emph {et~al.}(2020)\citenamefont
  {Takahashi}, \citenamefont {Kassa}, \citenamefont {Christoforou},\ and\
  \citenamefont {Keller}}]{takahashi_strong_2020}%
  \BibitemOpen
  \bibfield  {author} {\bibinfo {author} {\bibfnamefont {H.}~\bibnamefont
  {Takahashi}}, \bibinfo {author} {\bibfnamefont {E.}~\bibnamefont {Kassa}},
  \bibinfo {author} {\bibfnamefont {C.}~\bibnamefont {Christoforou}},\ and\
  \bibinfo {author} {\bibfnamefont {M.}~\bibnamefont {Keller}},\ }\bibfield
  {title} {\bibinfo {title} {Strong {Coupling} of a {Single} {Ion} to an
  {Optical} {Cavity}},\ }\href {https://doi.org/10.1103/PhysRevLett.124.013602}
  {\bibfield  {journal} {\bibinfo  {journal} {Phys. Rev. Lett.}\ }\textbf
  {\bibinfo {volume} {124}},\ \bibinfo {pages} {013602} (\bibinfo {year}
  {2020})}\BibitemShut {NoStop}%
\bibitem [{\citenamefont {Wang}\ \emph {et~al.}(2019)\citenamefont {Wang},
  \citenamefont {Kelkar}, \citenamefont {Martin-Cano}, \citenamefont
  {Rattenbacher}, \citenamefont {Shkarin}, \citenamefont {Utikal},
  \citenamefont {Götzinger},\ and\ \citenamefont
  {Sandoghdar}}]{wang_turning_2019}%
  \BibitemOpen
  \bibfield  {author} {\bibinfo {author} {\bibfnamefont {D.}~\bibnamefont
  {Wang}}, \bibinfo {author} {\bibfnamefont {H.}~\bibnamefont {Kelkar}},
  \bibinfo {author} {\bibfnamefont {D.}~\bibnamefont {Martin-Cano}}, \bibinfo
  {author} {\bibfnamefont {D.}~\bibnamefont {Rattenbacher}}, \bibinfo {author}
  {\bibfnamefont {A.}~\bibnamefont {Shkarin}}, \bibinfo {author} {\bibfnamefont
  {T.}~\bibnamefont {Utikal}}, \bibinfo {author} {\bibfnamefont
  {S.}~\bibnamefont {Götzinger}},\ and\ \bibinfo {author} {\bibfnamefont
  {V.}~\bibnamefont {Sandoghdar}},\ }\bibfield  {title} {{\selectlanguage
  {english}\bibinfo {title} {Turning a molecule into a coherent two-level
  quantum system}},\ }\href {https://doi.org/10.1038/s41567-019-0436-5}
  {\bibfield  {journal} {\bibinfo  {journal} {Nat. Phys.}\ }\textbf {\bibinfo
  {volume} {15}},\ \bibinfo {pages} {483} (\bibinfo {year} {2019})}\BibitemShut
  {NoStop}%
\bibitem [{\citenamefont {Pscherer}\ \emph {et~al.}(2021)\citenamefont
  {Pscherer}, \citenamefont {Meierhofer}, \citenamefont {Wang}, \citenamefont
  {Kelkar}, \citenamefont {Martín-Cano}, \citenamefont {Utikal}, \citenamefont
  {Götzinger},\ and\ \citenamefont
  {Sandoghdar}}]{pscherer_single-molecule_2021}%
  \BibitemOpen
  \bibfield  {author} {\bibinfo {author} {\bibfnamefont {A.}~\bibnamefont
  {Pscherer}}, \bibinfo {author} {\bibfnamefont {M.}~\bibnamefont
  {Meierhofer}}, \bibinfo {author} {\bibfnamefont {D.}~\bibnamefont {Wang}},
  \bibinfo {author} {\bibfnamefont {H.}~\bibnamefont {Kelkar}}, \bibinfo
  {author} {\bibfnamefont {D.}~\bibnamefont {Martín-Cano}}, \bibinfo {author}
  {\bibfnamefont {T.}~\bibnamefont {Utikal}}, \bibinfo {author} {\bibfnamefont
  {S.}~\bibnamefont {Götzinger}},\ and\ \bibinfo {author} {\bibfnamefont
  {V.}~\bibnamefont {Sandoghdar}},\ }\bibfield  {title} {\bibinfo {title}
  {Single-molecule vacuum {Rabi} splitting: four-wave mixing and optical
  switching at the single-photon level},\ }\href
  {https://journals.aps.org/prl/abstract/10.1103/PhysRevLett.127.133603}
  {\bibfield  {journal} {\bibinfo  {journal} {Phys. Rev. Lett.}\ }\textbf
  {\bibinfo {volume} {127}},\ \bibinfo {pages} {133603} (\bibinfo {year}
  {2021})}\BibitemShut {NoStop}%
\bibitem [{\citenamefont {Rakher}\ \emph {et~al.}(2009)\citenamefont {Rakher},
  \citenamefont {Stoltz}, \citenamefont {Coldren}, \citenamefont {Petroff},\
  and\ \citenamefont {Bouwmeester}}]{rakher_externally_2009}%
  \BibitemOpen
  \bibfield  {author} {\bibinfo {author} {\bibfnamefont {M.~T.}\ \bibnamefont
  {Rakher}}, \bibinfo {author} {\bibfnamefont {N.~G.}\ \bibnamefont {Stoltz}},
  \bibinfo {author} {\bibfnamefont {L.~A.}\ \bibnamefont {Coldren}}, \bibinfo
  {author} {\bibfnamefont {P.~M.}\ \bibnamefont {Petroff}},\ and\ \bibinfo
  {author} {\bibfnamefont {D.}~\bibnamefont {Bouwmeester}},\ }\bibfield
  {title} {\bibinfo {title} {Externally {Mode}-{Matched} {Cavity} {Quantum}
  {Electrodynamics} with {Charge}-{Tunable} {Quantum} {Dots}},\ }\href
  {https://doi.org/10.1103/PhysRevLett.102.097403} {\bibfield  {journal}
  {\bibinfo  {journal} {Phys. Rev. Lett.}\ }\textbf {\bibinfo {volume} {102}},\
  \bibinfo {pages} {097403} (\bibinfo {year} {2009})}\BibitemShut {NoStop}%
\bibitem [{\citenamefont {De~Santis}\ \emph {et~al.}(2017)\citenamefont
  {De~Santis}, \citenamefont {Antón}, \citenamefont {Reznychenko},
  \citenamefont {Somaschi}, \citenamefont {Coppola}, \citenamefont {Senellart},
  \citenamefont {Gómez}, \citenamefont {Lemaître}, \citenamefont {Sagnes},
  \citenamefont {White}, \citenamefont {Lanco}, \citenamefont {Auffèves},\
  and\ \citenamefont {Senellart}}]{de_santis_solid-state_2017}%
  \BibitemOpen
  \bibfield  {author} {\bibinfo {author} {\bibfnamefont {L.}~\bibnamefont
  {De~Santis}}, \bibinfo {author} {\bibfnamefont {C.}~\bibnamefont {Antón}},
  \bibinfo {author} {\bibfnamefont {B.}~\bibnamefont {Reznychenko}}, \bibinfo
  {author} {\bibfnamefont {N.}~\bibnamefont {Somaschi}}, \bibinfo {author}
  {\bibfnamefont {G.}~\bibnamefont {Coppola}}, \bibinfo {author} {\bibfnamefont
  {J.}~\bibnamefont {Senellart}}, \bibinfo {author} {\bibfnamefont
  {C.}~\bibnamefont {Gómez}}, \bibinfo {author} {\bibfnamefont
  {A.}~\bibnamefont {Lemaître}}, \bibinfo {author} {\bibfnamefont
  {I.}~\bibnamefont {Sagnes}}, \bibinfo {author} {\bibfnamefont {A.~G.}\
  \bibnamefont {White}}, \bibinfo {author} {\bibfnamefont {L.}~\bibnamefont
  {Lanco}}, \bibinfo {author} {\bibfnamefont {A.}~\bibnamefont {Auffèves}},\
  and\ \bibinfo {author} {\bibfnamefont {P.}~\bibnamefont {Senellart}},\
  }\bibfield  {title} {{\selectlanguage {english}\bibinfo {title} {A
  solid-state single-photon filter}},\ }\href
  {https://doi.org/10.1038/nnano.2017.85} {\bibfield  {journal} {\bibinfo
  {journal} {Nat. Nanotechnol.}\ }\textbf {\bibinfo {volume} {12}},\ \bibinfo
  {pages} {663} (\bibinfo {year} {2017})}\BibitemShut {NoStop}%
\bibitem [{\citenamefont {Najer}\ \emph {et~al.}(2019)\citenamefont {Najer},
  \citenamefont {S\"{o}llner}, \citenamefont {Sekatski}, \citenamefont
  {Dolique}, \citenamefont {L\"{o}bl}, \citenamefont {Riedel}, \citenamefont
  {Schott}, \citenamefont {Starosielec}, \citenamefont {Valentin},
  \citenamefont {Wieck}, \citenamefont {Sangouard}, \citenamefont {Ludwig},\
  and\ \citenamefont {Warburton}}]{Najer2019}%
  \BibitemOpen
  \bibfield  {author} {\bibinfo {author} {\bibfnamefont {D.}~\bibnamefont
  {Najer}}, \bibinfo {author} {\bibfnamefont {I.}~\bibnamefont {S\"{o}llner}},
  \bibinfo {author} {\bibfnamefont {P.}~\bibnamefont {Sekatski}}, \bibinfo
  {author} {\bibfnamefont {V.}~\bibnamefont {Dolique}}, \bibinfo {author}
  {\bibfnamefont {M.~C.}\ \bibnamefont {L\"{o}bl}}, \bibinfo {author}
  {\bibfnamefont {D.}~\bibnamefont {Riedel}}, \bibinfo {author} {\bibfnamefont
  {R.}~\bibnamefont {Schott}}, \bibinfo {author} {\bibfnamefont
  {S.}~\bibnamefont {Starosielec}}, \bibinfo {author} {\bibfnamefont {S.~R.}\
  \bibnamefont {Valentin}}, \bibinfo {author} {\bibfnamefont {A.~D.}\
  \bibnamefont {Wieck}}, \bibinfo {author} {\bibfnamefont {N.}~\bibnamefont
  {Sangouard}}, \bibinfo {author} {\bibfnamefont {A.}~\bibnamefont {Ludwig}},\
  and\ \bibinfo {author} {\bibfnamefont {R.~J.}\ \bibnamefont {Warburton}},\
  }\bibfield  {title} {\bibinfo {title} {A gated quantum dot strongly coupled
  to an optical microcavity},\ }\href
  {https://doi.org/10.1038/s41586-019-1709-y} {\bibfield  {journal} {\bibinfo
  {journal} {Nature}\ }\textbf {\bibinfo {volume} {575}},\ \bibinfo {pages}
  {622} (\bibinfo {year} {2019})}\BibitemShut {NoStop}%
\bibitem [{\citenamefont {Auffèves-Garnier}\ \emph {et~al.}(2007)\citenamefont
  {Auffèves-Garnier}, \citenamefont {Simon}, \citenamefont {Gérard},\ and\
  \citenamefont {Poizat}}]{auffeves-garnier_giant_2007}%
  \BibitemOpen
  \bibfield  {author} {\bibinfo {author} {\bibfnamefont {A.}~\bibnamefont
  {Auffèves-Garnier}}, \bibinfo {author} {\bibfnamefont {C.}~\bibnamefont
  {Simon}}, \bibinfo {author} {\bibfnamefont {J.-M.}\ \bibnamefont {Gérard}},\
  and\ \bibinfo {author} {\bibfnamefont {J.-P.}\ \bibnamefont {Poizat}},\
  }\bibfield  {title} {\bibinfo {title} {Giant optical nonlinearity induced by
  a single two-level system interacting with a cavity in the {Purcell}
  regime},\ }\href {https://doi.org/10.1103/PhysRevA.75.053823} {\bibfield
  {journal} {\bibinfo  {journal} {Phys. Rev. A}\ }\textbf {\bibinfo {volume}
  {75}},\ \bibinfo {pages} {053823} (\bibinfo {year} {2007})}\BibitemShut
  {NoStop}%
\bibitem [{\citenamefont {Kuhlmann}\ \emph {et~al.}(2013)\citenamefont
  {Kuhlmann}, \citenamefont {Houel}, \citenamefont {Ludwig}, \citenamefont
  {Greuter}, \citenamefont {Reuter}, \citenamefont {Wieck}, \citenamefont
  {Poggio},\ and\ \citenamefont {Warburton}}]{Kuhlmann2013}%
  \BibitemOpen
  \bibfield  {author} {\bibinfo {author} {\bibfnamefont {A.~V.}\ \bibnamefont
  {Kuhlmann}}, \bibinfo {author} {\bibfnamefont {J.}~\bibnamefont {Houel}},
  \bibinfo {author} {\bibfnamefont {A.}~\bibnamefont {Ludwig}}, \bibinfo
  {author} {\bibfnamefont {L.}~\bibnamefont {Greuter}}, \bibinfo {author}
  {\bibfnamefont {D.}~\bibnamefont {Reuter}}, \bibinfo {author} {\bibfnamefont
  {A.~D.}\ \bibnamefont {Wieck}}, \bibinfo {author} {\bibfnamefont
  {M.}~\bibnamefont {Poggio}},\ and\ \bibinfo {author} {\bibfnamefont {R.~J.}\
  \bibnamefont {Warburton}},\ }\bibfield  {title} {\bibinfo {title} {Charge
  noise and spin noise in a semiconductor quantum device},\ }\href
  {https://doi.org/10.1038/NPHYS2688} {\bibfield  {journal} {\bibinfo
  {journal} {Nat. Phys.}\ }\textbf {\bibinfo {volume} {9}},\ \bibinfo {pages}
  {570} (\bibinfo {year} {2013})}\BibitemShut {NoStop}%
\bibitem [{\citenamefont {Iles-Smith}\ \emph {et~al.}(2017)\citenamefont
  {Iles-Smith}, \citenamefont {McCutcheon}, \citenamefont {Nazir},\ and\
  \citenamefont {Mork}}]{IlesSmith2017}%
  \BibitemOpen
  \bibfield  {author} {\bibinfo {author} {\bibfnamefont {J.}~\bibnamefont
  {Iles-Smith}}, \bibinfo {author} {\bibfnamefont {D.~P.~S.}\ \bibnamefont
  {McCutcheon}}, \bibinfo {author} {\bibfnamefont {A.}~\bibnamefont {Nazir}},\
  and\ \bibinfo {author} {\bibfnamefont {J.}~\bibnamefont {Mork}},\ }\bibfield
  {title} {\bibinfo {title} {Phonon scattering inhibits simultaneous near-unity
  efficiency and indistinguishability in semiconductor single-photon sources},\
  }\href {https://doi.org/10.1038/NPHOTON.2017.101} {\bibfield  {journal}
  {\bibinfo  {journal} {Nat. Photon.}\ }\textbf {\bibinfo {volume} {11}},\
  \bibinfo {pages} {521} (\bibinfo {year} {2017})}\BibitemShut {NoStop}%
\bibitem [{\citenamefont {Shen}\ and\ \citenamefont
  {Fan}(2005)}]{shen_coherent_2005}%
  \BibitemOpen
  \bibfield  {author} {\bibinfo {author} {\bibfnamefont {J.~T.}\ \bibnamefont
  {Shen}}\ and\ \bibinfo {author} {\bibfnamefont {S.}~\bibnamefont {Fan}},\
  }\bibfield  {title} {{\selectlanguage {english}\bibinfo {title} {Coherent
  photon transport from spontaneous emission in one-dimensional waveguides}},\
  }\href {https://doi.org/10.1364/OL.30.002001} {\bibfield  {journal} {\bibinfo
   {journal} {Opt. Lett.}\ }\textbf {\bibinfo {volume} {30}},\ \bibinfo {pages}
  {2001} (\bibinfo {year} {2005})}\BibitemShut {NoStop}%
\bibitem [{\citenamefont {Shen}\ and\ \citenamefont
  {Fan}(2007{\natexlab{b}})}]{Shen2007}%
  \BibitemOpen
  \bibfield  {author} {\bibinfo {author} {\bibfnamefont {J.-T.}\ \bibnamefont
  {Shen}}\ and\ \bibinfo {author} {\bibfnamefont {S.}~\bibnamefont {Fan}},\
  }\bibfield  {title} {\bibinfo {title} {Strongly correlated two-photon
  transport in a one-dimensional waveguide coupled to a two-level system},\
  }\href {https://doi.org/10.1103/PhysRevLett.98.153003} {\bibfield  {journal}
  {\bibinfo  {journal} {Phys. Rev. Lett.}\ }\textbf {\bibinfo {volume} {98}},\
  \bibinfo {pages} {153003} (\bibinfo {year} {2007}{\natexlab{b}})}\BibitemShut
  {NoStop}%
\bibitem [{\citenamefont {Chang}\ \emph {et~al.}(2007)\citenamefont {Chang},
  \citenamefont {Sorensen}, \citenamefont {Demler},\ and\ \citenamefont
  {Lukin}}]{Chang2007}%
  \BibitemOpen
  \bibfield  {author} {\bibinfo {author} {\bibfnamefont {D.~E.}\ \bibnamefont
  {Chang}}, \bibinfo {author} {\bibfnamefont {A.~S.}\ \bibnamefont {Sorensen}},
  \bibinfo {author} {\bibfnamefont {E.~A.}\ \bibnamefont {Demler}},\ and\
  \bibinfo {author} {\bibfnamefont {M.~D.}\ \bibnamefont {Lukin}},\ }\bibfield
  {title} {\bibinfo {title} {A single-photon transistor using nanoscale surface
  plasmons},\ }\href {https://doi.org/10.1038/nphys708} {\bibfield  {journal}
  {\bibinfo  {journal} {Nat. Phys.}\ }\textbf {\bibinfo {volume} {3}},\
  \bibinfo {pages} {807} (\bibinfo {year} {2007})}\BibitemShut {NoStop}%
\bibitem [{\citenamefont {Javadi}\ \emph {et~al.}(2015)\citenamefont {Javadi},
  \citenamefont {S{\"o}llner}, \citenamefont {Arcari}, \citenamefont {Hansen},
  \citenamefont {Midolo}, \citenamefont {Mahmoodian}, \citenamefont
  {Kir{\v{s}}ansk{\.e}}, \citenamefont {Pregnolato}, \citenamefont {Lee},
  \citenamefont {Song}, \citenamefont {Stobbe},\ and\ \citenamefont
  {Lodahl}}]{Javadi2015NCOM}%
  \BibitemOpen
  \bibfield  {author} {\bibinfo {author} {\bibfnamefont {A.}~\bibnamefont
  {Javadi}}, \bibinfo {author} {\bibfnamefont {I.}~\bibnamefont {S{\"o}llner}},
  \bibinfo {author} {\bibfnamefont {M.}~\bibnamefont {Arcari}}, \bibinfo
  {author} {\bibfnamefont {S.~L.}\ \bibnamefont {Hansen}}, \bibinfo {author}
  {\bibfnamefont {L.}~\bibnamefont {Midolo}}, \bibinfo {author} {\bibfnamefont
  {S.}~\bibnamefont {Mahmoodian}}, \bibinfo {author} {\bibfnamefont
  {G.}~\bibnamefont {Kir{\v{s}}ansk{\.e}}}, \bibinfo {author} {\bibfnamefont
  {T.}~\bibnamefont {Pregnolato}}, \bibinfo {author} {\bibfnamefont
  {E.}~\bibnamefont {Lee}}, \bibinfo {author} {\bibfnamefont {J.}~\bibnamefont
  {Song}}, \bibinfo {author} {\bibfnamefont {S.}~\bibnamefont {Stobbe}},\ and\
  \bibinfo {author} {\bibfnamefont {P.}~\bibnamefont {Lodahl}},\ }\bibfield
  {title} {\bibinfo {title} {Single-photon non-linear optics with a quantum dot
  in a waveguide},\ }\href@noop {} {\bibfield  {journal} {\bibinfo  {journal}
  {Nat. Commun.}\ }\textbf {\bibinfo {volume} {6}},\ \bibinfo {pages} {8655}
  (\bibinfo {year} {2015})}\BibitemShut {NoStop}%
\bibitem [{\citenamefont {Rice}\ and\ \citenamefont
  {Carmichael}(1988)}]{rice_single-atom_1988}%
  \BibitemOpen
  \bibfield  {author} {\bibinfo {author} {\bibfnamefont {P.~R.}\ \bibnamefont
  {Rice}}\ and\ \bibinfo {author} {\bibfnamefont {H.~J.}\ \bibnamefont
  {Carmichael}},\ }\bibfield  {title} {\bibinfo {title} {Single-atom
  cavity-enhanced absorption. {I}. {Photon} statistics in the bad-cavity
  limit},\ }\href {https://doi.org/10.1109/3.974} {\bibfield  {journal}
  {\bibinfo  {journal} {IEEE J. Quantum Electron.}\ }\textbf {\bibinfo {volume}
  {24}},\ \bibinfo {pages} {1351} (\bibinfo {year} {1988})}\BibitemShut
  {NoStop}%
\bibitem [{sup()}]{suppmat}%
  \BibitemOpen
  \href@noop {} {}\bibinfo {note} {In the Supplemental Material we provide the
  full theoretical description used in the main text together with the
  connection the simplified scenarios and the construction of the reduced
  cavity states (Rice-Carmichael states). We also include a table with all the
  parameters used in each figure in the main text for the theory curves. In the
  SM, we include Refs.~\cite{GardinerCollet1985, Auffeves-Garnier2007PRA,
  rice_single-atom_1988, Breuer&Petruccione2002, Carmichael2008,
  WallsMilburn2008}}\BibitemShut {NoStop}%
\bibitem [{\citenamefont {Barbour}\ \emph {et~al.}(2011)\citenamefont
  {Barbour}, \citenamefont {Dalgarno}, \citenamefont {Curran}, \citenamefont
  {Nowak}, \citenamefont {Baker}, \citenamefont {Hall}, \citenamefont {Stoltz},
  \citenamefont {Petroff},\ and\ \citenamefont {Warburton}}]{Barbour2011}%
  \BibitemOpen
  \bibfield  {author} {\bibinfo {author} {\bibfnamefont {R.~J.}\ \bibnamefont
  {Barbour}}, \bibinfo {author} {\bibfnamefont {P.~A.}\ \bibnamefont
  {Dalgarno}}, \bibinfo {author} {\bibfnamefont {A.}~\bibnamefont {Curran}},
  \bibinfo {author} {\bibfnamefont {K.~M.}\ \bibnamefont {Nowak}}, \bibinfo
  {author} {\bibfnamefont {H.~J.}\ \bibnamefont {Baker}}, \bibinfo {author}
  {\bibfnamefont {D.~R.}\ \bibnamefont {Hall}}, \bibinfo {author}
  {\bibfnamefont {N.~G.}\ \bibnamefont {Stoltz}}, \bibinfo {author}
  {\bibfnamefont {P.~M.}\ \bibnamefont {Petroff}},\ and\ \bibinfo {author}
  {\bibfnamefont {R.~J.}\ \bibnamefont {Warburton}},\ }\bibfield  {title}
  {\bibinfo {title} {A tunable microcavity},\ }\href
  {https://doi.org/10.1063/1.3632057} {\bibfield  {journal} {\bibinfo
  {journal} {J. Appl. Phys.}\ }\textbf {\bibinfo {volume} {110}},\ \bibinfo
  {pages} {053107} (\bibinfo {year} {2011})}\BibitemShut {NoStop}%
\bibitem [{\citenamefont {Tomm}\ \emph {et~al.}(2021)\citenamefont {Tomm},
  \citenamefont {Javadi}, \citenamefont {Antoniadis}, \citenamefont {Najer},
  \citenamefont {L\"obl}, \citenamefont {Korsch}, \citenamefont {Schott},
  \citenamefont {Valentin}, \citenamefont {Wieck}, \citenamefont {Ludwig},\
  and\ \citenamefont {Warburton}}]{Tomm2021}%
  \BibitemOpen
  \bibfield  {author} {\bibinfo {author} {\bibfnamefont {N.}~\bibnamefont
  {Tomm}}, \bibinfo {author} {\bibfnamefont {A.}~\bibnamefont {Javadi}},
  \bibinfo {author} {\bibfnamefont {N.~O.}\ \bibnamefont {Antoniadis}},
  \bibinfo {author} {\bibfnamefont {D.}~\bibnamefont {Najer}}, \bibinfo
  {author} {\bibfnamefont {M.~C.}\ \bibnamefont {L\"obl}}, \bibinfo {author}
  {\bibfnamefont {A.~R.}\ \bibnamefont {Korsch}}, \bibinfo {author}
  {\bibfnamefont {R.}~\bibnamefont {Schott}}, \bibinfo {author} {\bibfnamefont
  {S.~R.}\ \bibnamefont {Valentin}}, \bibinfo {author} {\bibfnamefont {A.~D.}\
  \bibnamefont {Wieck}}, \bibinfo {author} {\bibfnamefont {A.}~\bibnamefont
  {Ludwig}},\ and\ \bibinfo {author} {\bibfnamefont {R.~J.}\ \bibnamefont
  {Warburton}},\ }\bibfield  {title} {\bibinfo {title} {A bright and fast
  source of coherent single photons},\ }\href
  {https://www.nature.com/articles/s41565-020-00831-x} {\bibfield  {journal}
  {\bibinfo  {journal} {Nat. Nanotechnol.}\ }\textbf {\bibinfo {volume} {16}},\
  \bibinfo {pages} {399} (\bibinfo {year} {2021})}\BibitemShut {NoStop}%
\bibitem [{\citenamefont {Tomm}\ \emph {et~al.}(2023)\citenamefont {Tomm},
  \citenamefont {Mahmoodian}, \citenamefont {Antoniadis}, \citenamefont
  {Schott}, \citenamefont {Valentin}, \citenamefont {Wieck}, \citenamefont
  {Ludwig}, \citenamefont {Javadi},\ and\ \citenamefont
  {Warburton}}]{Tomm_direct2022}%
  \BibitemOpen
  \bibfield  {author} {\bibinfo {author} {\bibfnamefont {N.}~\bibnamefont
  {Tomm}}, \bibinfo {author} {\bibfnamefont {S.}~\bibnamefont {Mahmoodian}},
  \bibinfo {author} {\bibfnamefont {N.~O.}\ \bibnamefont {Antoniadis}},
  \bibinfo {author} {\bibfnamefont {R.}~\bibnamefont {Schott}}, \bibinfo
  {author} {\bibfnamefont {S.~R.}\ \bibnamefont {Valentin}}, \bibinfo {author}
  {\bibfnamefont {A.~D.}\ \bibnamefont {Wieck}}, \bibinfo {author}
  {\bibfnamefont {A.}~\bibnamefont {Ludwig}}, \bibinfo {author} {\bibfnamefont
  {A.}~\bibnamefont {Javadi}},\ and\ \bibinfo {author} {\bibfnamefont {R.~J.}\
  \bibnamefont {Warburton}},\ }\bibfield  {title} {\bibinfo {title} {Direct
  observation of photon bound states using a single artificial atom},\ }\href
  {https://www.nature.com/articles/s41567-023-01997-6} {\bibfield  {journal}
  {\bibinfo  {journal} {Nat. Phys.}\ }\textbf {\bibinfo {volume} {19}},\
  \bibinfo {pages} {857} (\bibinfo {year} {2023})}\BibitemShut {NoStop}%
\bibitem [{Note1()}]{Note1}%
  \BibitemOpen
  \bibinfo {note} {The theory predicts $P_{\protect \textrm {sat}}=1.5$ nW; the
  mismatch is attributed to losses \cite {suppmat}.}\BibitemShut {Stop}%
\bibitem [{\citenamefont {Carmichael}(2008)}]{Carmichael2008}%
  \BibitemOpen
  \bibfield  {author} {\bibinfo {author} {\bibfnamefont {H.}~\bibnamefont
  {Carmichael}},\ }\href
  {https://doi.org/https://doi.org/10.1007/978-3-540-71320-3} {\emph {\bibinfo
  {title} {Statistical Methods in Quantum Optics 2}}}\ (\bibinfo  {publisher}
  {Springer Berlin Heidelberg},\ \bibinfo {address} {Berlin, Heidelberg},\
  \bibinfo {year} {2008})\BibitemShut {NoStop}%
\bibitem [{\citenamefont {Carmichael}\ \emph {et~al.}(1989)\citenamefont
  {Carmichael}, \citenamefont {Brecha}, \citenamefont {Raizen}, \citenamefont
  {Kimble},\ and\ \citenamefont {Rice}}]{Carmichael1989PRA}%
  \BibitemOpen
  \bibfield  {author} {\bibinfo {author} {\bibfnamefont {H.~J.}\ \bibnamefont
  {Carmichael}}, \bibinfo {author} {\bibfnamefont {R.~J.}\ \bibnamefont
  {Brecha}}, \bibinfo {author} {\bibfnamefont {M.~G.}\ \bibnamefont {Raizen}},
  \bibinfo {author} {\bibfnamefont {H.~J.}\ \bibnamefont {Kimble}},\ and\
  \bibinfo {author} {\bibfnamefont {P.~R.}\ \bibnamefont {Rice}},\ }\bibfield
  {title} {\bibinfo {title} {Subnatural linewidth averaging for coupled atomic
  and cavity-mode oscillators},\ }\href
  {https://journals.aps.org/pra/abstract/10.1103/PhysRevA.40.5516} {\bibfield
  {journal} {\bibinfo  {journal} {Phys. Rev. A}\ }\textbf {\bibinfo {volume}
  {40}},\ \bibinfo {pages} {5516} (\bibinfo {year} {1989})}\BibitemShut
  {NoStop}%
\bibitem [{\citenamefont {Lodahl}\ \emph {et~al.}(2017)\citenamefont {Lodahl},
  \citenamefont {Mahmoodian}, \citenamefont {Stobbe}, \citenamefont
  {Rauschenbeutel}, \citenamefont {Schneeweiss}, \citenamefont {Volz},
  \citenamefont {Pichler},\ and\ \citenamefont {Zoller}}]{lodahl_chiral_2017}%
  \BibitemOpen
  \bibfield  {author} {\bibinfo {author} {\bibfnamefont {P.}~\bibnamefont
  {Lodahl}}, \bibinfo {author} {\bibfnamefont {S.}~\bibnamefont {Mahmoodian}},
  \bibinfo {author} {\bibfnamefont {S.}~\bibnamefont {Stobbe}}, \bibinfo
  {author} {\bibfnamefont {A.}~\bibnamefont {Rauschenbeutel}}, \bibinfo
  {author} {\bibfnamefont {P.}~\bibnamefont {Schneeweiss}}, \bibinfo {author}
  {\bibfnamefont {J.}~\bibnamefont {Volz}}, \bibinfo {author} {\bibfnamefont
  {H.}~\bibnamefont {Pichler}},\ and\ \bibinfo {author} {\bibfnamefont
  {P.}~\bibnamefont {Zoller}},\ }\bibfield  {title} {{\selectlanguage
  {english}\bibinfo {title} {Chiral quantum optics}},\ }\href
  {https://doi.org/10.1038/nature21037} {\bibfield  {journal} {\bibinfo
  {journal} {Nature}\ }\textbf {\bibinfo {volume} {541}},\ \bibinfo {pages}
  {473} (\bibinfo {year} {2017})}\BibitemShut {NoStop}%
\bibitem [{\citenamefont {Antoniadis}\ \emph {et~al.}(2022)\citenamefont
  {Antoniadis}, \citenamefont {Tomm}, \citenamefont {Jakubczyk}, \citenamefont
  {Schott}, \citenamefont {Valentin}, \citenamefont {Wieck}, \citenamefont
  {Ludwig}, \citenamefont {Warburton},\ and\ \citenamefont
  {Javadi}}]{antoniadis_chiral_2022}%
  \BibitemOpen
  \bibfield  {author} {\bibinfo {author} {\bibfnamefont {N.~O.}\ \bibnamefont
  {Antoniadis}}, \bibinfo {author} {\bibfnamefont {N.}~\bibnamefont {Tomm}},
  \bibinfo {author} {\bibfnamefont {T.}~\bibnamefont {Jakubczyk}}, \bibinfo
  {author} {\bibfnamefont {R.}~\bibnamefont {Schott}}, \bibinfo {author}
  {\bibfnamefont {S.~R.}\ \bibnamefont {Valentin}}, \bibinfo {author}
  {\bibfnamefont {A.~D.}\ \bibnamefont {Wieck}}, \bibinfo {author}
  {\bibfnamefont {A.}~\bibnamefont {Ludwig}}, \bibinfo {author} {\bibfnamefont
  {R.~J.}\ \bibnamefont {Warburton}},\ and\ \bibinfo {author} {\bibfnamefont
  {A.}~\bibnamefont {Javadi}},\ }\bibfield  {title} {{\selectlanguage
  {english}\bibinfo {title} {A chiral one-dimensional atom using a quantum dot
  in an open microcavity}},\ }\href
  {https://doi.org/10.1038/s41534-022-00545-z} {\bibfield  {journal} {\bibinfo
  {journal} {npj Quantum Inf.}\ }\textbf {\bibinfo {volume} {8}},\ \bibinfo
  {pages} {27} (\bibinfo {year} {2022})}\BibitemShut {NoStop}%
\bibitem [{\citenamefont {Mahmoodian}\ \emph {et~al.}(2020)\citenamefont
  {Mahmoodian}, \citenamefont {Calaj\'o}, \citenamefont {Chang}, \citenamefont
  {Hammerer},\ and\ \citenamefont {S\o{}rensen}}]{Mahmoodian2020}%
  \BibitemOpen
  \bibfield  {author} {\bibinfo {author} {\bibfnamefont {S.}~\bibnamefont
  {Mahmoodian}}, \bibinfo {author} {\bibfnamefont {G.}~\bibnamefont
  {Calaj\'o}}, \bibinfo {author} {\bibfnamefont {D.~E.}\ \bibnamefont {Chang}},
  \bibinfo {author} {\bibfnamefont {K.}~\bibnamefont {Hammerer}},\ and\
  \bibinfo {author} {\bibfnamefont {A.~S.}\ \bibnamefont {S\o{}rensen}},\
  }\bibfield  {title} {\bibinfo {title} {Dynamics of many-body photon bound
  states in chiral waveguide {QED}},\ }\href
  {https://doi.org/10.1103/PhysRevX.10.031011} {\bibfield  {journal} {\bibinfo
  {journal} {Phys. Rev. X}\ }\textbf {\bibinfo {volume} {10}},\ \bibinfo
  {pages} {031011} (\bibinfo {year} {2020})}\BibitemShut {NoStop}%
\bibitem [{\citenamefont {Stiesdal}\ \emph {et~al.}(2018)\citenamefont
  {Stiesdal}, \citenamefont {Kumlin}, \citenamefont {Kleinbeck}, \citenamefont
  {Lunt}, \citenamefont {Braun}, \citenamefont {Paris-Mandoki}, \citenamefont
  {Tresp}, \citenamefont {B\"uchler},\ and\ \citenamefont
  {Hofferberth}}]{Stiesdal2018}%
  \BibitemOpen
  \bibfield  {author} {\bibinfo {author} {\bibfnamefont {N.}~\bibnamefont
  {Stiesdal}}, \bibinfo {author} {\bibfnamefont {J.}~\bibnamefont {Kumlin}},
  \bibinfo {author} {\bibfnamefont {K.}~\bibnamefont {Kleinbeck}}, \bibinfo
  {author} {\bibfnamefont {P.}~\bibnamefont {Lunt}}, \bibinfo {author}
  {\bibfnamefont {C.}~\bibnamefont {Braun}}, \bibinfo {author} {\bibfnamefont
  {A.}~\bibnamefont {Paris-Mandoki}}, \bibinfo {author} {\bibfnamefont
  {C.}~\bibnamefont {Tresp}}, \bibinfo {author} {\bibfnamefont {H.~P.}\
  \bibnamefont {B\"uchler}},\ and\ \bibinfo {author} {\bibfnamefont
  {S.}~\bibnamefont {Hofferberth}},\ }\bibfield  {title} {\bibinfo {title}
  {Observation of three-body correlations for photons coupled to a {R}ydberg
  superatom},\ }\href {https://doi.org/10.1103/PhysRevLett.121.103601}
  {\bibfield  {journal} {\bibinfo  {journal} {Phys. Rev. Lett.}\ }\textbf
  {\bibinfo {volume} {121}},\ \bibinfo {pages} {103601} (\bibinfo {year}
  {2018})}\BibitemShut {NoStop}%
\bibitem [{\citenamefont {Liang}\ \emph {et~al.}(2018)\citenamefont {Liang},
  \citenamefont {Venkatramani}, \citenamefont {Cantu}, \citenamefont
  {Nicholson}, \citenamefont {Gullans}, \citenamefont {Gorshkov}, \citenamefont
  {Thompson}, \citenamefont {Chin}, \citenamefont {Lukin},\ and\ \citenamefont
  {Vuletić}}]{Liang2018}%
  \BibitemOpen
  \bibfield  {author} {\bibinfo {author} {\bibfnamefont {Q.-Y.}\ \bibnamefont
  {Liang}}, \bibinfo {author} {\bibfnamefont {A.~V.}\ \bibnamefont
  {Venkatramani}}, \bibinfo {author} {\bibfnamefont {S.~H.}\ \bibnamefont
  {Cantu}}, \bibinfo {author} {\bibfnamefont {T.~L.}\ \bibnamefont
  {Nicholson}}, \bibinfo {author} {\bibfnamefont {M.~J.}\ \bibnamefont
  {Gullans}}, \bibinfo {author} {\bibfnamefont {A.~V.}\ \bibnamefont
  {Gorshkov}}, \bibinfo {author} {\bibfnamefont {J.~D.}\ \bibnamefont
  {Thompson}}, \bibinfo {author} {\bibfnamefont {C.}~\bibnamefont {Chin}},
  \bibinfo {author} {\bibfnamefont {M.~D.}\ \bibnamefont {Lukin}},\ and\
  \bibinfo {author} {\bibfnamefont {V.}~\bibnamefont {Vuletić}},\ }\bibfield
  {title} {\bibinfo {title} {Observation of three-photon bound states in a
  quantum nonlinear medium},\ }\href {https://doi.org/10.1126/science.aao7293}
  {\bibfield  {journal} {\bibinfo  {journal} {Science}\ }\textbf {\bibinfo
  {volume} {359}},\ \bibinfo {pages} {783} (\bibinfo {year}
  {2018})}\BibitemShut {NoStop}%
\bibitem [{\citenamefont {Noh}\ and\ \citenamefont
  {Angelakis}(2016)}]{noh_quantum_2016}%
  \BibitemOpen
  \bibfield  {author} {\bibinfo {author} {\bibfnamefont {C.}~\bibnamefont
  {Noh}}\ and\ \bibinfo {author} {\bibfnamefont {D.~G.}\ \bibnamefont
  {Angelakis}},\ }\bibfield  {title} {{\selectlanguage {english}\bibinfo
  {title} {Quantum simulations and many-body physics with light}},\ }\href
  {https://doi.org/10.1088/0034-4885/80/1/016401} {\bibfield  {journal}
  {\bibinfo  {journal} {Rep. Prog. Phys.}\ }\textbf {\bibinfo {volume} {80}},\
  \bibinfo {pages} {016401} (\bibinfo {year} {2016})}\BibitemShut {NoStop}%
\bibitem [{\citenamefont {Ralph}\ \emph {et~al.}(2015)\citenamefont {Ralph},
  \citenamefont {Söllner}, \citenamefont {Mahmoodian}, \citenamefont {White},\
  and\ \citenamefont {Lodahl}}]{ralph_photon_2015}%
  \BibitemOpen
  \bibfield  {author} {\bibinfo {author} {\bibfnamefont {T.~C.}\ \bibnamefont
  {Ralph}}, \bibinfo {author} {\bibfnamefont {I.}~\bibnamefont {Söllner}},
  \bibinfo {author} {\bibfnamefont {S.}~\bibnamefont {Mahmoodian}}, \bibinfo
  {author} {\bibfnamefont {A.}~\bibnamefont {White}},\ and\ \bibinfo {author}
  {\bibfnamefont {P.}~\bibnamefont {Lodahl}},\ }\bibfield  {title} {\bibinfo
  {title} {Photon {Sorting}, {Efficient} {Bell} {Measurements}, and a
  {Deterministic} {Controlled}-{Z} {Gate} {Using} a {Passive} {Two}-{Level}
  {Nonlinearity}},\ }\href {https://doi.org/10.1103/PhysRevLett.114.173603}
  {\bibfield  {journal} {\bibinfo  {journal} {Phys. Rev. Lett.}\ }\textbf
  {\bibinfo {volume} {114}},\ \bibinfo {pages} {173603} (\bibinfo {year}
  {2015})}\BibitemShut {NoStop}%
\bibitem [{\citenamefont {Yang}\ \emph {et~al.}(2022)\citenamefont {Yang},
  \citenamefont {Lund}, \citenamefont {Pohl}, \citenamefont {Lodahl},\ and\
  \citenamefont {Mølmer}}]{yang_deterministic_2022}%
  \BibitemOpen
  \bibfield  {author} {\bibinfo {author} {\bibfnamefont {F.}~\bibnamefont
  {Yang}}, \bibinfo {author} {\bibfnamefont {M.~M.}\ \bibnamefont {Lund}},
  \bibinfo {author} {\bibfnamefont {T.}~\bibnamefont {Pohl}}, \bibinfo {author}
  {\bibfnamefont {P.}~\bibnamefont {Lodahl}},\ and\ \bibinfo {author}
  {\bibfnamefont {K.}~\bibnamefont {Mølmer}},\ }\bibfield  {title} {\bibinfo
  {title} {Deterministic {Photon} {Sorting} in {Waveguide} {QED} {Systems}},\
  }\href {https://doi.org/10.1103/PhysRevLett.128.213603} {\bibfield  {journal}
  {\bibinfo  {journal} {Phys. Rev. Lett.}\ }\textbf {\bibinfo {volume} {128}},\
  \bibinfo {pages} {213603} (\bibinfo {year} {2022})}\BibitemShut {NoStop}%
\bibitem [{\citenamefont {Witthaut}\ \emph {et~al.}(2012)\citenamefont
  {Witthaut}, \citenamefont {Lukin},\ and\ \citenamefont
  {S{\o}rensen}}]{Witthaut2012}%
  \BibitemOpen
  \bibfield  {author} {\bibinfo {author} {\bibfnamefont {D.}~\bibnamefont
  {Witthaut}}, \bibinfo {author} {\bibfnamefont {M.~D.}\ \bibnamefont
  {Lukin}},\ and\ \bibinfo {author} {\bibfnamefont {A.~S.}\ \bibnamefont
  {S{\o}rensen}},\ }\bibfield  {title} {\bibinfo {title} {Photon sorters and
  {QND} detectors using single photon emitters},\ }\href
  {https://doi.org/10.1209/0295-5075/97/50007} {\bibfield  {journal} {\bibinfo
  {journal} {{EPL}}\ }\textbf {\bibinfo {volume} {97}},\ \bibinfo {pages}
  {50007} (\bibinfo {year} {2012})}\BibitemShut {NoStop}%
\bibitem [{\citenamefont {Shomroni}\ \emph {et~al.}(2014)\citenamefont
  {Shomroni}, \citenamefont {Rosenblum}, \citenamefont {Lovsky}, \citenamefont
  {Bechler}, \citenamefont {Guendelman},\ and\ \citenamefont
  {Dayan}}]{shomroni_all-optical_2014}%
  \BibitemOpen
  \bibfield  {author} {\bibinfo {author} {\bibfnamefont {I.}~\bibnamefont
  {Shomroni}}, \bibinfo {author} {\bibfnamefont {S.}~\bibnamefont {Rosenblum}},
  \bibinfo {author} {\bibfnamefont {Y.}~\bibnamefont {Lovsky}}, \bibinfo
  {author} {\bibfnamefont {O.}~\bibnamefont {Bechler}}, \bibinfo {author}
  {\bibfnamefont {G.}~\bibnamefont {Guendelman}},\ and\ \bibinfo {author}
  {\bibfnamefont {B.}~\bibnamefont {Dayan}},\ }\bibfield  {title}
  {{\selectlanguage {english}\bibinfo {title} {All-optical routing of single
  photons by a one-atom switch controlled by a single photon}},\ }\href
  {https://doi.org/10.1126/science.1254699} {\bibfield  {journal} {\bibinfo
  {journal} {Science}\ }\textbf {\bibinfo {volume} {345}},\ \bibinfo {pages}
  {903} (\bibinfo {year} {2014})}\BibitemShut {NoStop}%
\bibitem [{\citenamefont {Chen}\ \emph {et~al.}(2021)\citenamefont {Chen},
  \citenamefont {Zhou}, \citenamefont {Shen}, \citenamefont {Ku},\ and\
  \citenamefont {Steel}}]{chen_two-photon_2021}%
  \BibitemOpen
  \bibfield  {author} {\bibinfo {author} {\bibfnamefont {Z.}~\bibnamefont
  {Chen}}, \bibinfo {author} {\bibfnamefont {Y.}~\bibnamefont {Zhou}}, \bibinfo
  {author} {\bibfnamefont {J.-T.}\ \bibnamefont {Shen}}, \bibinfo {author}
  {\bibfnamefont {P.-C.}\ \bibnamefont {Ku}},\ and\ \bibinfo {author}
  {\bibfnamefont {D.}~\bibnamefont {Steel}},\ }\bibfield  {title} {\bibinfo
  {title} {Two-photon controlled-phase gates enabled by photonic dimers},\
  }\href {https://doi.org/10.1103/PhysRevA.103.052610} {\bibfield  {journal}
  {\bibinfo  {journal} {Phys. Rev. A}\ }\textbf {\bibinfo {volume} {103}},\
  \bibinfo {pages} {052610} (\bibinfo {year} {2021})}\BibitemShut {NoStop}%
\bibitem [{\citenamefont {Brod}\ and\ \citenamefont
  {Combes}(2016)}]{brod_passive_2016}%
  \BibitemOpen
  \bibfield  {author} {\bibinfo {author} {\bibfnamefont {D.~J.}\ \bibnamefont
  {Brod}}\ and\ \bibinfo {author} {\bibfnamefont {J.}~\bibnamefont {Combes}},\
  }\bibfield  {title} {\bibinfo {title} {Passive {CPHASE} {Gate} via
  {Cross}-{Kerr} {Nonlinearities}},\ }\href
  {https://doi.org/10.1103/PhysRevLett.117.080502} {\bibfield  {journal}
  {\bibinfo  {journal} {Phys. Rev. Lett.}\ }\textbf {\bibinfo {volume} {117}},\
  \bibinfo {pages} {080502} (\bibinfo {year} {2016})},\ \bibinfo {note}
  {publisher American Physical Society}\BibitemShut {NoStop}%
\bibitem [{\citenamefont {Gardiner}\ and\ \citenamefont
  {Collett}(1985)}]{GardinerCollet1985}%
  \BibitemOpen
  \bibfield  {author} {\bibinfo {author} {\bibfnamefont {C.~W.}\ \bibnamefont
  {Gardiner}}\ and\ \bibinfo {author} {\bibfnamefont {M.~J.}\ \bibnamefont
  {Collett}},\ }\bibfield  {title} {\bibinfo {title} {Input and output in
  damped quantum systems: Quantum stochastic differential equations and the
  master equation},\ }\href {https://doi.org/10.1103/PhysRevA.31.3761}
  {\bibfield  {journal} {\bibinfo  {journal} {Phys. Rev. A}\ }\textbf {\bibinfo
  {volume} {31}},\ \bibinfo {pages} {3761} (\bibinfo {year}
  {1985})}\BibitemShut {NoStop}%
\bibitem [{\citenamefont {Auff\`eves-Garnier}\ \emph
  {et~al.}(2007)\citenamefont {Auff\`eves-Garnier}, \citenamefont {Simon},
  \citenamefont {G\'erard},\ and\ \citenamefont
  {Poizat}}]{Auffeves-Garnier2007PRA}%
  \BibitemOpen
  \bibfield  {author} {\bibinfo {author} {\bibfnamefont {A.}~\bibnamefont
  {Auff\`eves-Garnier}}, \bibinfo {author} {\bibfnamefont {C.}~\bibnamefont
  {Simon}}, \bibinfo {author} {\bibfnamefont {J.-M.}\ \bibnamefont
  {G\'erard}},\ and\ \bibinfo {author} {\bibfnamefont {J.-P.}\ \bibnamefont
  {Poizat}},\ }\bibfield  {title} {\bibinfo {title} {Giant optical nonlinearity
  induced by a single two-level system interacting with a cavity in the purcell
  regime},\ }\href {https://doi.org/10.1103/PhysRevA.75.053823} {\bibfield
  {journal} {\bibinfo  {journal} {Phys. Rev. A}\ }\textbf {\bibinfo {volume}
  {75}},\ \bibinfo {pages} {053823} (\bibinfo {year} {2007})}\BibitemShut
  {NoStop}%
\bibitem [{\citenamefont {Breuer}\ and\ \citenamefont
  {Petruccione}(2002)}]{Breuer&Petruccione2002}%
  \BibitemOpen
  \bibfield  {author} {\bibinfo {author} {\bibfnamefont {H.-P.}\ \bibnamefont
  {Breuer}}\ and\ \bibinfo {author} {\bibfnamefont {F.}~\bibnamefont
  {Petruccione}},\ }\href@noop {} {\emph {\bibinfo {title} {The Theory of Open
  Quantum Systems}}}\ (\bibinfo  {publisher} {Oxford University Press},\
  \bibinfo {year} {2002})\BibitemShut {NoStop}%
\bibitem [{\citenamefont {Walls}\ and\ \citenamefont
  {Milburn}(2008)}]{WallsMilburn2008}%
  \BibitemOpen
  \bibfield  {author} {\bibinfo {author} {\bibfnamefont {D.~F.}\ \bibnamefont
  {Walls}}\ and\ \bibinfo {author} {\bibfnamefont {G.~J.}\ \bibnamefont
  {Milburn}},\ }\href
  {https://doi.org/https://doi.org/10.1007/978-3-540-28574-8} {\emph {\bibinfo
  {title} {Quantum Optics}}},\ \bibinfo {edition} {2nd}\ ed.\ (\bibinfo
  {publisher} {Springer Verlag Berlin},\ \bibinfo {year} {2008})\BibitemShut
  {NoStop}%
\end{thebibliography}%


\begin{thebibliography}{6}%
\makeatletter
\providecommand \@ifxundefined [1]{%
 \@ifx{#1\undefined}
}%
\providecommand \@ifnum [1]{%
 \ifnum #1\expandafter \@firstoftwo
 \else \expandafter \@secondoftwo
 \fi
}%
\providecommand \@ifx [1]{%
 \ifx #1\expandafter \@firstoftwo
 \else \expandafter \@secondoftwo
 \fi
}%
\providecommand \natexlab [1]{#1}%
\providecommand \enquote  [1]{``#1''}%
\providecommand \bibnamefont  [1]{#1}%
\providecommand \bibfnamefont [1]{#1}%
\providecommand \citenamefont [1]{#1}%
\providecommand \href@noop [0]{\@secondoftwo}%
\providecommand \href [0]{\begingroup \@sanitize@url \@href}%
\providecommand \@href[1]{\@@startlink{#1}\@@href}%
\providecommand \@@href[1]{\endgroup#1\@@endlink}%
\providecommand \@sanitize@url [0]{\catcode `\\12\catcode `\$12\catcode
  `\&12\catcode `\#12\catcode `\^12\catcode `\_12\catcode `\%12\relax}%
\providecommand \@@startlink[1]{}%
\providecommand \@@endlink[0]{}%
\providecommand \url  [0]{\begingroup\@sanitize@url \@url }%
\providecommand \@url [1]{\endgroup\@href {#1}{\urlprefix }}%
\providecommand \urlprefix  [0]{URL }%
\providecommand \Eprint [0]{\href }%
\providecommand \doibase [0]{https://doi.org/}%
\providecommand \selectlanguage [0]{\@gobble}%
\providecommand \bibinfo  [0]{\@secondoftwo}%
\providecommand \bibfield  [0]{\@secondoftwo}%
\providecommand \translation [1]{[#1]}%
\providecommand \BibitemOpen [0]{}%
\providecommand \bibitemStop [0]{}%
\providecommand \bibitemNoStop [0]{.\EOS\space}%
\providecommand \EOS [0]{\spacefactor3000\relax}%
\providecommand \BibitemShut  [1]{\csname bibitem#1\endcsname}%
\let\auto@bib@innerbib\@empty
\bibitem [{\citenamefont {Gardiner}\ and\ \citenamefont
  {Collett}(1985)}]{GardinerCollet1985}%
  \BibitemOpen
  \bibfield  {author} {\bibinfo {author} {\bibfnamefont {C.~W.}\ \bibnamefont
  {Gardiner}}\ and\ \bibinfo {author} {\bibfnamefont {M.~J.}\ \bibnamefont
  {Collett}},\ }\bibfield  {title} {\bibinfo {title} {Input and output in
  damped quantum systems: Quantum stochastic differential equations and the
  master equation},\ }\href {https://doi.org/10.1103/PhysRevA.31.3761}
  {\bibfield  {journal} {\bibinfo  {journal} {Phys. Rev. A}\ }\textbf {\bibinfo
  {volume} {31}},\ \bibinfo {pages} {3761} (\bibinfo {year}
  {1985})}\BibitemShut {NoStop}%
\bibitem [{\citenamefont {Walls}\ and\ \citenamefont
  {Milburn}(2008)}]{WallsMilburn2008}%
  \BibitemOpen
  \bibfield  {author} {\bibinfo {author} {\bibfnamefont {D.~F.}\ \bibnamefont
  {Walls}}\ and\ \bibinfo {author} {\bibfnamefont {G.~J.}\ \bibnamefont
  {Milburn}},\ }\href
  {https://doi.org/https://doi.org/10.1007/978-3-540-28574-8} {\emph {\bibinfo
  {title} {Quantum Optics}}},\ \bibinfo {edition} {2nd}\ ed.\ (\bibinfo
  {publisher} {Springer Verlag Berlin},\ \bibinfo {year} {2008})\BibitemShut
  {NoStop}%
\bibitem [{\citenamefont {Breuer}\ and\ \citenamefont
  {Petruccione}(2002)}]{Breuer&Petruccione2002}%
  \BibitemOpen
  \bibfield  {author} {\bibinfo {author} {\bibfnamefont {H.-P.}\ \bibnamefont
  {Breuer}}\ and\ \bibinfo {author} {\bibfnamefont {F.}~\bibnamefont
  {Petruccione}},\ }\href@noop {} {\emph {\bibinfo {title} {The Theory of Open
  Quantum Systems}}}\ (\bibinfo  {publisher} {Oxford University Press},\
  \bibinfo {year} {2002})\BibitemShut {NoStop}%
\bibitem [{\citenamefont {Rice}\ and\ \citenamefont
  {Carmichael}(1988)}]{rice_single-atom_1988}%
  \BibitemOpen
  \bibfield  {author} {\bibinfo {author} {\bibfnamefont {P.~R.}\ \bibnamefont
  {Rice}}\ and\ \bibinfo {author} {\bibfnamefont {H.~J.}\ \bibnamefont
  {Carmichael}},\ }\bibfield  {title} {\bibinfo {title} {Single-atom
  cavity-enhanced absorption. {I}. {Photon} statistics in the bad-cavity
  limit},\ }\href {https://doi.org/10.1109/3.974} {\bibfield  {journal}
  {\bibinfo  {journal} {IEEE J. Quantum Electron.}\ }\textbf {\bibinfo {volume}
  {24}},\ \bibinfo {pages} {1351} (\bibinfo {year} {1988})},\ \bibinfo {note}
  {conference Name: IEEE Journal of Quantum Electronics}\BibitemShut {NoStop}%
\bibitem [{\citenamefont {Auffèves-Garnier}\ \emph {et~al.}(2007)\citenamefont
  {Auffèves-Garnier}, \citenamefont {Simon}, \citenamefont {Gérard},\ and\
  \citenamefont {Poizat}}]{auffeves-garnier_giant_2007}%
  \BibitemOpen
  \bibfield  {author} {\bibinfo {author} {\bibfnamefont {A.}~\bibnamefont
  {Auffèves-Garnier}}, \bibinfo {author} {\bibfnamefont {C.}~\bibnamefont
  {Simon}}, \bibinfo {author} {\bibfnamefont {J.-M.}\ \bibnamefont {Gérard}},\
  and\ \bibinfo {author} {\bibfnamefont {J.-P.}\ \bibnamefont {Poizat}},\
  }\bibfield  {title} {\bibinfo {title} {Giant optical nonlinearity induced by
  a single two-level system interacting with a cavity in the {Purcell}
  regime},\ }\href {https://doi.org/10.1103/PhysRevA.75.053823} {\bibfield
  {journal} {\bibinfo  {journal} {Phys. Rev. A}\ }\textbf {\bibinfo {volume}
  {75}},\ \bibinfo {pages} {053823} (\bibinfo {year} {2007})},\ \bibinfo {note}
  {publisher American Physical Society}\BibitemShut {NoStop}%
\bibitem [{\citenamefont {Carmichael}(2008)}]{Carmichael2008-cQED2}%
  \BibitemOpen
  \bibfield  {author} {\bibinfo {author} {\bibfnamefont {H.}~\bibnamefont
  {Carmichael}},\ }\bibinfo {title} {Statistical methods in quantum optics 2:
  Non-classical fields. cavity qed ii: Quantum fluctuations},\ in\ \href
  {https://doi.org/10.1007/978-3-540-71320-3_8} {\emph {\bibinfo {booktitle}
  {Statistical Methods in Quantum Optics 2: Non-Classical Fields}}}\ (\bibinfo
  {publisher} {Springer Berlin Heidelberg},\ \bibinfo {address} {Berlin,
  Heidelberg},\ \bibinfo {year} {2008})\ pp.\ \bibinfo {pages}
  {335--400}\BibitemShut {NoStop}%
\end{thebibliography}%
\end{document}


\title{Supplemental Material: Realisation of a Coherent and Efficient One-Dimensional Atom}
\author{Natasha Tomm}
\thanks{These authors contributed equally}
\affiliation{Department of Physics, University of Basel, Klingelbergstrasse 82, CH-4056 Basel, Switzerland}
\author{Nadia O. Antoniadis}
\thanks{These authors contributed equally}
\affiliation{Department of Physics, University of Basel, Klingelbergstrasse 82, CH-4056 Basel, Switzerland}
\author{Marcelo Janovitch}
\thanks{These authors contributed equally}
\affiliation{Department of Physics, University of Basel, Klingelbergstrasse 82, CH-4056 Basel, Switzerland}
\author{Matteo Brunelli}
\affiliation{Department of Physics, University of Basel, Klingelbergstrasse 82, CH-4056 Basel, Switzerland}
\author{R\"{u}diger Schott} 
\affiliation{Lehrstuhl f\"{u}r Angewandte Festk\"{o}rperphysik, Ruhr-Universit\"{a}t Bochum, D-44780 Bochum, Germany}
\author{Sascha R. Valentin}
\affiliation{Lehrstuhl f\"{u}r Angewandte Festk\"{o}rperphysik, Ruhr-Universit\"{a}t Bochum, D-44780 Bochum, Germany}
\author{Andreas D. Wieck}
\affiliation{Lehrstuhl f\"{u}r Angewandte Festk\"{o}rperphysik, Ruhr-Universit\"{a}t Bochum, D-44780 Bochum, Germany}
\author{Arne Ludwig}
\affiliation{Lehrstuhl f\"{u}r Angewandte Festk\"{o}rperphysik, Ruhr-Universit\"{a}t Bochum, D-44780 Bochum, Germany}
\author{Patrick P. Potts}
\affiliation{Department of Physics, University of Basel, Klingelbergstrasse 82, CH-4056 Basel, Switzerland}
\author{Alisa Javadi} 
\altaffiliation{Present address: School of Electrical and Computer Engineering, Department of Physics and Astronomy, The University of Oklahoma, 110 West Boyd Street, OK 73019, USA} 
\affiliation{Department of Physics, University of Basel, Klingelbergstrasse 82, CH-4056 Basel, Switzerland}
\author{Richard J. Warburton}
\email[To whom correspondence should be addressed:]{\\ m.janovitch@unibas.ch, nadia.antoniadis@unibas.ch}
\affiliation{Department of Physics, University of Basel, Klingelbergstrasse 82, CH-4056 Basel, Switzerland}

\date{\today}	
\maketitle

Here, we describe the theoretical model used to compute transmission, reflection and intensity correlations in both transmission mode and reflection mode configurations. Whenever it is necessary (e.g. for $g^{(2)}$--functions) we denote the reflection mode with ``$\leftarrow$'' to distinguish from the transmission mode. We start by describing the cavity-dot (system) dynamics through a Lindblad master equation (LME). We then express the measured quantities using input-output theory in terms of system quantities, computed through the LME. Then, we employ the adiabatic elimination and perturbation theory in the drive strength, $\epsilon$, to solve for steady-state and obtain correlation functions. Further, we construct the Rice-Carmichael states for each experiment and connect them with the intensity-correlations. Finally, we also include a table with all the parameters used for the theoretical curves presented in the main text.

\tableofcontents
 
\vspace{1cm}

\section{General theory}
\subsection{Lindblad master equation}
We consider a LME of the form,
\begin{align}\label{eq:lme}
    \dv{\rhos}{t} = \mathcal{L}\rhos = -i[\ham,\rhos] + \kappa\sum_{\mu=\text{H,V}} D[\hat{a}_\mu]\rhos +  \gamma\sum_{j=a,b}D[\Sj]\rhos,
\end{align}
where $\hat{\rho}_S$ is the density matrix of the cavity and quantum dot (QD) $ \hbar=1$, $D[\hat{L}]\hat{\rho} = \hat{L}\hat{\rho} \hat{L}^\dagger - 1/2(\hat{L}^\dagger \hat{L} \hat{\rho} + \hat{\rho} \hat{L}^\dagger \hat{L}) $, $\dyad{g}{j}$ are the quantum dot (QD) transition operators between the ground state ($\ket{g}$) and the excited states ($\ket{j}$, with $j=a,b$), $\kappa$ the dissipation rate of the cavity and $\gamma$ the loss rate of the QD. The Hamiltonian in a rotating frame relative to the laser frequency, $\hat{H} = \hat{H}_0 + \hat{H}_\text{int} + \hat{H}_\text{drive}$ is composed of
\begin{align}
    &\ham_0 = -\sum_{j=a, b} \Delta\omega_j \dyad{j} - \sum_{\mu=H,V} \Delta\omega_\mu \hat{a}^\dagger_\mu \hat{a}_\mu,\label{eq:ham0}\\
    &\ham_\text{int} = \sum_{j,\mu} g_{\mu,j} \qty(\hat{a}^\dagger_\mu \Sj + \Sjd \hat{a}_\mu),\label{eq:hamint}\\
    &\ham_\text{drive} = -\sum_{\mu=\text{H,V}}\sqrt{\kappa} \epsilon_\mu \qty(\hat{a}^\dagger_\mu + \hat{a}_\mu),\label{eq:eff}
\end{align}
with $~\Delta \omega_{\mu/j} = \omega_\text{laser} - \omega_{\mu/j}$ detunings with respect to the input laser frequency, $g_{\mu,j}$ the coupling constant between the QD level $j$ and the cavity mode $\mu$ and $\epsilon_\mu$ the photon flux from the laser (in units of $1/\sqrt{\text{time}}$). 

In general, as represented in Fig.~1\,(b) of the main text, the QD transitions are not aligned with the cavity's polarisation axes, however, the  QD transitions are mutually orthogonal and we assume that they couple symmetrically to the light modes due to the symmetry of the setup, i.e., $g_{H,a} = g_{V,b} = \sqrt{g}\cos\theta$, $g_{H,b} = g_{V,a} = -\sqrt{g}\sin\theta$.  The angle $\theta$ is the relative angle between the QD polarisation axes and the cavity polarisation axes and $g$ denotes the overall coupling strength.
We then introduce the Purcell-enhanced decay rate
\begin{equation}
    \Gamma = \frac{2g^2}{\kappa},
\end{equation}
the Purcell factor
\begin{equation}
F_\text{P} =  \frac{2\Gamma}{\gamma} = \frac{4g^2}{\kappa\gamma},
\end{equation}
and the $\beta$-factor
\begin{equation}
\label{eq:beta}
    \beta = \frac{F_{\text P}}{1+F_{\text P}}.
\end{equation}

We also adopt the convention $\epsilon^2 = \epsilon_\text{H}^2 + \epsilon_\text{V}^2$. In the transmission mode measurements, this gives $\epsilon_\mu = \epsilon /\sqrt{2}$ and, in the reflection mode measurements, $\epsilon_\text{H} = \epsilon,~\epsilon_\text{V} = 0$.

Equation~\eqref{eq:lme} allows computation of the steady-state, $\hat{\rho}_\sss$, and thereby the average of any observable $\E{\hat{\mathcal{O}}}_\sss$. We postpone the explicit calculation of these and show first the formal connection with the quantities experimentally accessed and discussed in the main text.

\subsection{Input-output theory: reflection and transmission functions}

The LME~\eqref{eq:lme} does not explicitly include the field outside the cavity, which is detected.
The connection can be established by the input-output formalism, noting that Eq.~\eqref{eq:lme} can be derived from a quantum Langevin equation~\cite{GardinerCollet1985}.
In this case, the drive is captured by input modes $\bin{\mu},~\mu = \text{H, V}$ which obey
\begin{align}
    &[\bin{\mu}(\tau), \bin{\nu}^\dagger(\tau')] = \delta_{\mu, \nu} \delta(\tau-\tau'),\\
    &\E{\bin{\mu}} =-i\epsilon_\mu,
\end{align}
for a continuous-wave laser, in a frame rotating with the laser angular frequency $\omega_\text{laser}$, where we fixed the phase, $-i$, consistent with Eq.~\eqref{eq:eff}. 
The experimentally accessible quantities are the output modes, given by the input-output relation (IOR)
\begin{align}\label{eq:ior}
    \bout{\mu} = \bin{\mu} + \sqrt{\kappa} \hat{a}_\mu,
\end{align}
where $\bin{\mu}$ and $\bout{\mu}$ have units $1/\sqrt{\text{time}}$ and cavity modes are adimensional. 

The IOR allows us to define the transmission and reflection amplitudes. For transmission mode measurements, we input light in the mode $\bin{\text{P}} =(\bin{\text{H}} + \bin{\text{V}})/\sqrt{2}$ and collect in the orthogonal mode, $\bout{\text{M}}=(\bout{\text{H}}-\bout{\text{V}})/\sqrt{2}$ and define
\begin{align}
    t &= \frac{\E{\bout{\text{M}}}}{\E{\bin{\text{P}}}} = i \frac{\sqrt{\kappa}}{\epsilon} \frac{\E{\hat{a}_\text{H} - \hat{a}_\text{V}}_\sss}{\sqrt{2}}\label{eq:tT}\\
     r &=  \frac{\E{\bout{\text{P}}}}{\E{\bin{\text{P}}}} =1+ i \frac{\sqrt{\kappa}}{\epsilon} \frac{\E{\hat{a}_\text{H} + \hat{a}_\text{V}}_\sss}{\sqrt{2}}.
\end{align}
For the reflection mode experiments, we input in $\bin{\text{H}}$ and collect in  $\bout{\text{V}}$
\begin{align}
    t_\leftarrow &= \frac{\E{\bout{\text{H}}}}{\E{\bin{\text{V}}}} = i \frac{\sqrt{\kappa}}{\epsilon} \E{\hat{a}_\text{H}}_\sss,\\
     r_\leftarrow &= \frac{\E{\bout{\text{H}}}}{\E{\bin{\text{H}}}} =1+ i \frac{\sqrt{\kappa}}{\epsilon} \E{\hat{a}_\text{H}
     }_\sss.\label{eq:rBR}
\end{align}
In  the main text, we discuss measurements of $T=|t|^2$ (transmission mode) and $R_\leftarrow=|r_\leftarrow|^2$ (reflection mode). We note that, due to losses, $|t_{(\leftarrow)}|^2 + |r_{(\leftarrow)}|^2\neq 1$. To derive analytical expressions for Eqs.~(\ref{eq:tT}-\ref{eq:rBR}), we need to obtain steady-state solutions $\E{\hat{a}_\mu}_\sss$, which can be computed directly using the Lindblad master equation~\eqref{eq:lme}.

\subsection{Connection between the full models and the single-mode Jaynes-Cummings model}
\label{sec:fulljc}

We can establish a formal connection between the JC model mentioned in the main text and the full model which captures the experimental data using input-output theory.

First, we discuss the connection between the transmission mode experiment and a two-level system in a two-sided single-mode cavity. In the full model the IORs~\eqref{eq:ior} give
\begin{subequations}\label{eq:IORsTM}
\begin{align}
    \bout{\text{P}} &= \bin{\text{P}} + \sqrt{\kappa}\hat{a}_\text{P},\\
    \bout{\text{M}} &= \bin{\text{M}} + \sqrt{\kappa}\hat{a}_\text{M},
\end{align}
\end{subequations}
where $\hat{a}_\text{P/M}=\qty(\hat{a}_\text{H} \pm \hat{a}_\text{V})/{\sqrt{2}}$.
When $|\delta_\text{cav}|/\kappa \to \infty$, only $\text{H}$ polarised light enters the cavity and for all normal-ordered quantities we can set $\hat{a}_\text{V}=0$. Equations~\eqref{eq:IORsTM} then reduce to
\begin{align}
    \bout{\text{P}} &= \bin{\text{P}} + \sqrt{\frac{\kappa}{2}}\hat{a}_\text{H},\\
    \bout{\text{M}} &= \bin{\text{M}} + \sqrt{\frac{\kappa}{2}}\hat{a}_\text{H}.
\end{align}
The above equations are the same as for a single-cavity mode coupled to a two-sided cavity, see for instance Ref.~\cite{WallsMilburn2008}. Finally, if we further assume that $\theta=0$, only the $\dyad{g}{a}$ transition couples to H, and we have effectively a two-level system.

The connection between reflection mode experiment and a two-level system in a one-sided, single-mode cavity is much simpler: since the input is H-polarised, we only need QD transition alignment, $\theta =0$, so that both V--mode and $\dyad{g}{b}$--transition are decoupled from the dynamics.
\subsection{Intensity correlations}

We now address the intensity correlations. For the transmission-mode measurements, we have
\begin{align}
  g^{(2)}(\tau) = \frac{\E{\bout{\text{M}}^\dagger(0) \bout{\text{M}}^\dagger(\tau)\bout{\text{M}}(\tau) \bout{\text{M}}(0)}}{\E{\bout{\text{M}}^\dagger(0)\bout{\text{M}}(0)}^2} =\frac{\E{\hat{a}_\text{M}^\dagger \hat{a}_\text{M}^\dagger(\tau) \hat{a}_\text{M}(\tau) \hat{a}_\text{M} }_\sss}{\E{\hat{a}^\dagger_\text{M} \hat{a}_\text{M}}_\sss^2},
\end{align}
where we used the IOR~\eqref{eq:ior}. The above equation means that the transmission mode intensity correlations capture only the light emitted by the corresponding \textit{cavity} mode. For the reflection mode, however, interference with the laser background (input) contributes,
\begin{align}\label{eq:formalg2BR}
     g^{(2)}_\leftarrow(\tau) &= \frac{\E{\bout{\text{H}}^\dagger(0) \bout{\text{H}}^\dagger(\tau)\bout{\text{H}}(\tau) \bout{\text{H}}(0)}}{\E{\bout{\text{H}}^\dagger(0)\bout{\text{H}}(0)}^2}\\
     &= \frac{\expval{\qty(i\epsilon+\sqrt{\kappa}\hat{a}_{\text{H}}^\dagger)\qty(i\epsilon + \sqrt{\kappa}\hat{a}_{\text{H}}^\dagger(\tau))\qty(-i\epsilon + \sqrt{\kappa}\hat{a}_\text{H}(\tau))\qty(-i\epsilon + \sqrt{\kappa}\hat{a}_\text{H})}_\sss}{\expval{\qty(i\epsilon+\sqrt{\kappa}\hat{a}_\text{H}^\dagger)\qty(-i\epsilon+\sqrt{\kappa}\hat{a}_\text{H})}_\sss^2},
\end{align}
where we again used the IOR~\eqref{eq:ior}, the fact that the input field is in a coherent state, and standard commutation relations of input-output theory (see Eqs.(2.23-2.26) of Ref.~Ref.~\cite{GardinerCollet1985}).
We can see from the expressions of the intensity-correlations that their calculation relies only on system quantities. In terms of the Lindblad dynamics~\eqref{eq:lme}, a generic two-time correlation function can be computed using the quantum regression theorem
~\cite{Breuer&Petruccione2002}. For steady-state correlation functions, this requires finding a set of steady-state averages, that serve as initial conditions, and obtaining the propagator. This can all be obtained by solving the Heisenberg dynamics of this open quantum system, as follows.

\subsection{Effective QD dynamics through adiabatic elimination of the cavity}
We now work the equations of motion for observables of interest. It is more convenient to work in the Heisenberg picture, and thus with the adjoint of Eq.~\eqref{eq:lme},
\begin{align}\label{eq:lmed}
    \dv{\hat{\mathcal{O}}}{t} = \bar{\mathcal{L}}\hat{\mathcal{O}} =  i[\ham,\hat{\mathcal{O}}]
    + \kappa\sum_{\mu=H,V} \bar{D}[\hat{a}_\mu]\hat{\mathcal{O}} +  \gamma\sum_{j=a,b}\bar{D}[\Sj]\hat{\mathcal{O}},
\end{align}
where, for a generic operator $\hat{\mathcal{O}}$, $\bar{D}[\hat{L}]\hat{\mathcal{O}} = \hat{L}^\dagger\hat{\mathcal{O}}\hat{L} - 1/2(\hat{L}^\dagger \hat{L} \hat{\mathcal{O}} + \hat{\mathcal{O}}\hat{L}^\dagger \hat{L})$. 
We obtain for the cavity operators
\begin{align}
    \dv{\hat{a}_\text{H}}{t}&=\bar{\mathcal{L}}\hat{a}_\text{H} =\qty( i \Delta\omega_\text{H} - \frac{\kappa}{2})\hat{a}_\text{H} +i\sqrt{\frac{\kappa}{2}}\qty[\epsilon_\text{H} -\sqrt{\Gamma}\qty(\Sa \cos\theta - \Sb \sin\theta)],\\
    \dv{\hat{a}_\text{V}}{t}&=\bar{\mathcal{L}}\hat{a}_\text{V} =\qty(i \Delta\omega_\text{V} - \frac{\kappa}{2})\hat{a}_\text{V}+i\sqrt{\frac{\kappa}{2}}\qty[\epsilon_\text{V} +\sqrt{\Gamma}(\Sa \sin \theta - \Sb \cos \theta)]. 
\end{align}
 Whenever $\theta=0$, the QD transition $\dyad{a}{g}$ is aligned with the H-mode and the QD transition $\dyad{b}{g}$ with the V-mode; in the experimental scenario, we have $\theta = 25.1^\circ$. 
 
 We now adiabatically eliminate the cavity modes, which is suitable for typical experimental parameters: in GHz, $\kappa/(2\pi) = 28, \gamma/(2\pi) = 0.3, \epsilon^2/(2\pi) \approx 0.0001$. 
Since the cavity dissipates fast (large $\kappa$, also called the bad cavity limit), we may set ${\rm d}{a}_\mu/{\rm dt}=0$ in the above equations of motions which results in~Refs.~\cite{rice_single-atom_1988, auffeves-garnier_giant_2007},
    \begin{subequations}\label{eq:adb}
        \begin{align}
        \hat{a}_\text{H} = 2i t_\text{H}\frac{\epsilon_\text{H}}{\sqrt{\kappa}}\qty[1 - \sqrt{\frac{\Gamma}{2}}\frac{1}{\epsilon_\text{H}}\qty(\Sa \cos \theta - \Sb \sin \theta) ],\\
        \hat{a}_\text{V} = 2i t_\text{V}\frac{\epsilon_\text{V}}{\sqrt{\kappa}}\qty[1 + \sqrt{\frac{\Gamma}{2}}\frac{1}{\epsilon_\text{V}}\qty(\Sa \sin \theta - \Sb \cos\theta) ],
        \end{align}
    \end{subequations}
where we also introduced the bare cavity transmission amplitudes $t_\mu = (-2i\Delta \omega_\mu/\kappa + 1)^{-1}$ , which are complex and satisfy $|t_\mu|^2\leq 1$. The effect of  $\Re t_\mu$ is to reduce the Purcell-enhanced decay while $\Im t_\mu$ causes a shift in the QD transition frequencies proportional to the Purcell factor [e.g., see Eq.~\eqref{eq:g2brsimpl}]. Now, we can use these to obtain effective three-level system dynamics by substituting in the QD's equations of motion
\begin{subequations}\label{eq:eoms}
\begin{align}
    \dv{t}\dyad{a}=&-\qty[\gamma+ 2\Gamma(\cos^2\theta \Re t_\text{H} + \sin^2\theta\Re t_\text{V})] \dyad{a} + \Gamma \cos \theta \sin\theta\qty[ (t_\text{H} + t_\text{V}) \Sab+ \text{h.c.}]\\ 
    &+ \sqrt{2 \Gamma}\qty[ (\cos\theta \epsilon_\text{H} t_\text{H}^* - \sin \theta \epsilon_\text{V} t_\text{V}^*) \Sa + \text{h.c.}] \nonumber\\
    \dv{t}\Sa =& -\qty[\frac{\gamma}{2} +i \Delta \omega_a + \Gamma \qty(t_\text{H} \cos\theta^2 + t_\text{V} \sin\theta^2)]\Sa - \Gamma \cos\theta \sin\theta \qty(t_\text{H} + t_\text{V}) \Sb \\
    &+ \sqrt{2 \Gamma}\qty(\epsilon_\text{H} t_\text{H} \sin\theta - \epsilon_\text{V} t_\text{V} \cos\theta) \qty[2 \dyad{a} + \dyad{b}] - \sqrt{2 \Gamma}\qty(\epsilon_\text{H} t_\text{H} \cos\theta + \epsilon_\text{V} t_\text{V} \sin\theta)\nonumber ,\\
    \dv{t}\Sab=&-\qty[\gamma + i (\Delta \omega_a - \Delta \omega_b) + \Gamma((t_\text{V}+t_\text{H}^*) \cos\theta^2 + (t_\text{H} + t_\text{V}^*)\sin\theta^2)] \Sab\\
    &- \Gamma \cos\theta \sin\theta\qty[(t_\text{H} + t_\text{V}) \dyad{a} + (t_\text{H}^* + t_\text{V}^*) \dyad{b}]\nonumber\\
    &-\sqrt{2 \Gamma}(\epsilon_\text{H} t_\text{H}^* \cos\theta + \epsilon_\text{V} t_\text{V}^* \sin\theta) \Sb +\sqrt{2 \Gamma}(\epsilon_\text{H} t_\text{H} \sin\theta - \epsilon_\text{V} t_\text{V} \cos\theta) \Sad \nonumber .
\end{align}
\end{subequations}
 Either by taking the Hermitian-conjugate or switching $(a, \text{H})\leftrightarrow (b, \text{V})$ in the above, we obtain the equations of motion for the remaining operators. Thereof, we obtain closed set of differential equations describing the evolution of the QD
\begin{align}
 &\dv{\Vec{\sigma}}{t}  = G\Vec{\sigma} + \Vec{f}\\
 &\Vec{\sigma}= \qty(\dyad{a}, \dyad{b}, \Sa, \Sb, \Sab, \Sad, \Sbd, \Sba)^T.
 \end{align}
 \subsection{Weak drive perturbation  theory (small $\epsilon$)}
 In addition to the adiabatic elimination, we now consider a weak drive limit, i.e., small $\epsilon$. 
 To this end, we write the equations of motion~\eqref{eq:eoms} as
 \begin{align}
     &\dv{\Vec{\sigma}}{t}  = (G_0 + W)\Vec{\sigma} + \Vec{f},
 \end{align}
 where $G_0$ is independent of the drive strength while $W$ and $\vec{f}$ are linear in $\epsilon_\mu$.
 This allows us to look for a perturbative solution for the steady-state averages $\E{\Vec{\sigma}}_{\infty}$ and the propagator in the weak drive limit. Most experimental data is taken in this limit, corresponding to low power. We now outline how this is formally done and defer specific results for transmission and reflection mode measurements to below.

 For the steady-state averages, we seek a perturbative expansion for the inverse of $G$, since
 \begin{align}
     \E{\Vec{\sigma}}_\sss = -G^{-1} \Vec{f}.
 \end{align}
 We consider the Neumann series up to the first order,
 \begin{align}\label{eq:neumann}
     G^{-1} = G^{-1}_0 + G^{-1}_0W G^{-1}_0 .
 \end{align}
 This expansion is enough to account for the experimental data for low power, resulting in transmission and reflection which are independent of $\epsilon$.
 
We are also interested in the intensity correlations. To this end, we consider deviations from the steady-state averages, which are described by the differential equation, see~Ref.~\cite{Breuer&Petruccione2002}
 \begin{align}
     &\dv{\delta\Vec{\sigma}}{t}  = (G_0 + W)\delta\Vec{\sigma},
 \end{align}
$\delta \Vec{\sigma}=\Vec{\sigma} - \langle\Vec{\sigma}\rangle_\sss$. To obtain the leading order expansion in $\epsilon$ for the $g^{(2)}$--function, we used the Dyson series up to second order
\begin{align}\label{eq:dyson}
    e^{t G} = e^{tG_0} + \int_0^t \dd s~ e^{(t-s)G_0} W e^{sG_0}+ \int_0^t \int_0^s\dd s \dd s'~ e^{(t-s)G_0} W e^{(s-s')G_0} W e^{s'G_0}.
\end{align}
\section{Transmission mode experiment}
\subsection{Transmission function}\label{ss:transmission}
\begin{figure}[t]
    \centering
    \includegraphics[width = \textwidth]{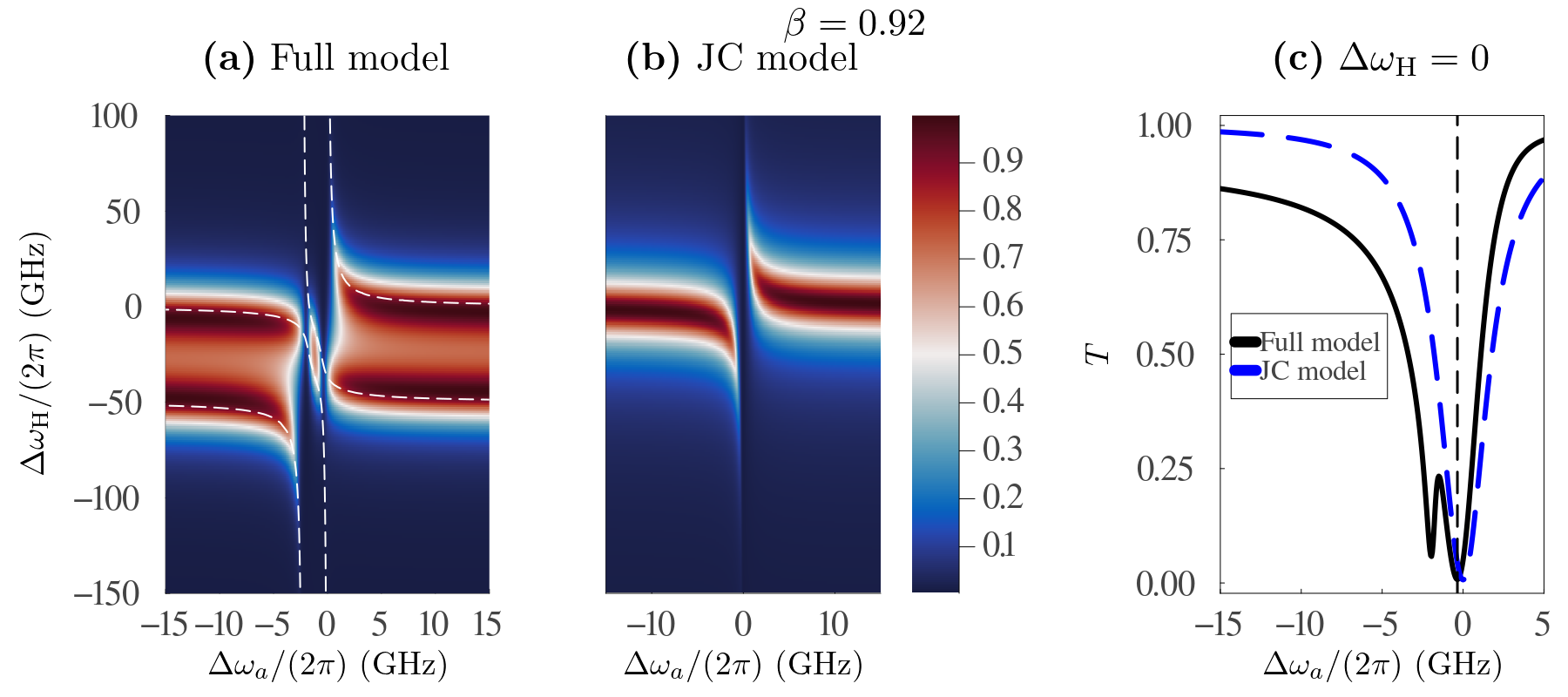} 
    \caption{Theoretical results for transmission, to be compared with Fig.~1\,(c) in the main text. In GHz, we set $g/(2\pi) = 4.8,~ \kappa/(2\pi)=28,~\delta_\text{QD}/(2\pi)=2.3,~ \gamma/(2\pi)=0.3$. \textbf{(a)} Full model with $\theta = 25.1^\circ$ and $\delta_\text{cav}/(2\pi)=50$. The white dashed lines are the points in which the gap between the ground and the first four excited states of $\hat{H}_0 + \hat{H}_{\rm int}$ [Eqs.~\eqref{eq:ham0} and \eqref{eq:hamint}] close. \textbf{(b)} JC model obtained by taking $\theta=0$ and $|\delta_{\rm cav}|/\kappa \to \infty$.~\textbf{(c)} Transmission function on cavity resonance $\Delta \omega_\text{H}=0$ for both models.}
    \label{fig:fig1_SI}
\end{figure}

Combining the formal expression for the transmission amplitude~\eqref{eq:tT} with the adiabatically eliminated modes~\eqref{eq:adb} we obtain the transmission amplitude in terms of the QD steady-state quantities,
\begin{align}
t= -(t_\text{H}-t_\text{V}) + \frac{\sqrt{\Gamma}}{\epsilon}\qty(\E{\Sa}_\sss( t_\text{H}\cos\theta + t_\text{V} \sin \theta) - \E{\Sb}_\sss( t_\text{V} \cos\theta + t_\text{H} \sin \theta)).
\end{align}
Where we note that for the transmission mode $\epsilon_\text{H} = \epsilon_\text{V} = \epsilon/\sqrt{2}$. If we compute the steady-state averages to first order in $\epsilon$ using Eq.~\eqref{eq:neumann} we obtain,
\begin{align} \label{eq:tTresult}
   &t =  -(t_\text{H}-t_\text{V}) + \frac{\beta \mathcal{N}_1 + \frac{\beta^2}{2}\mathcal{N}_2}{(\gamma - 2 i\Delta \omega_a)(\gamma - 2i\Delta\omega_b) + \beta \mathcal{D}_1+ \frac{\beta^2}{2}\mathcal{D}_2},
\end{align}
where we wrote the transmission amplitude in terms of $\beta$, c.f.~Eq.~\eqref{eq:beta}, and in terms of the coefficients
\begin{align}
   &\mathcal{N}_1 = \gamma\qty[(t_\text{H}^2 - t_\text{V}^2)(\gamma- i(\Delta \omega_a + \Delta \omega_b)) + i(t_\text{H}^2 + t_\text{V}^2))(\Delta \omega_a -\Delta \omega_b) \cos2\theta],\\
   &\mathcal{N}_2 = \gamma\Big[-2i(t_\text{H}^2 + t_\text{V}^2)(\Delta\omega_a - \Delta \omega_b) \cos 2\theta\\
   &~~~~~~~~+ (t_\text{H} - t_\text{V}) (-2 t_\text{H} (\gamma - i (\Delta \omega_a+\Delta \omega_b))-2 t_\text{V} (\gamma - i (\Delta \omega_a+\Delta \omega_b))+ \gamma t_\text{H} t_\text{V}(1+\cos 4\theta))\Big],\nonumber\\
   &\mathcal{D}_1 = \qty(t_\text{H} + t_\text{V} - 2) \gamma^2 + 8\Delta \omega_a \Delta \omega_b -(t_\text{H} + t_\text{V} -4)(\Delta \omega_a + \Delta \omega_b)\gamma + (t_\text{H} -t_\text{V})(\Delta \omega_a -\Delta \omega_b) \gamma \cos 2\theta),\\
   &\mathcal{D}_2 =\gamma^2(2 - 2t_\text{H} - 2t_\text{V} + t_\text{H} t_\text{V}) - 8 \Delta \omega_a \Delta \omega_b + 2i\gamma (-2 + t_\text{H} + t_\text{V}) (\Delta \omega_a + \Delta \omega_b)\\
   &~~~~~~~~- 2i\gamma (t_\text{H} - t_\text{V})(\Delta \omega_a - \Delta \omega_b) \cos 2\theta + \gamma^2t_\text{H} t_\text{V} \cos4\theta .\nonumber
\end{align}
The transmission function corresponding to the data in the main text is given by $T = |t|^2$. We observe that for perfect coupling efficiency, $\beta = 1$, $t =0$; we highlight that this idealised scenario can only be achieved asymptotically for $F_\text{P}\to\infty$. In the experiment, most relevant parameters can be tuned to a great extent, as discussed in the main text. 

We now discuss relevant limits. The simplest scenario to compare with Eq.~\eqref{eq:tTresult} is a single transition (two-level system) in a double-sided one-mode cavity (JC model). As discussed in Sec.~\ref{sec:fulljc}, our model reduces to the JC model for $|\delta_\text{cav}|/\kappa \to \infty$ and $\theta=0$, resulting in the transmission
\begin{align}
    t\to - t_{\rm H} + \frac{t^2_{\rm H}\beta}{1 -(1-t_{\rm H})\beta -2i (1-\beta) \Delta \omega_a/\gamma}. 
\end{align}
In this case, we can clearly see that for $\beta < 1$ the effect of dissipation can be minimised by bringing the QD in resonance with the laser, $\Delta \omega_a = 0$, and we obtain $T = (1-\beta)^2$. Also, in this limit, the splitting between QD levels is irrelevant since the second excited state does not participate in the dynamics.

 We now move to a scenario that is closer to the experiment, assuming resonance with the H-mode and that the QD transitions are perfectly aligned, $\Delta \omega_\text{H} =0,~\theta = 0$. In this case, we obtain
\begin{align}
     t\to -1 + t_\text{V} + \beta\qty( \frac{1}{1 - 2i (\beta - 1) \Delta \omega_a/\gamma} - \frac{t_\text{V}^2}{1 - 2i (\beta - 1) \Delta \omega_b/\gamma+ (t_\text{V} -1)\beta}),
\end{align}
which reduces to the previous limit for $t_\text{V} =0$. The above expression showcases that the detuned second cavity mode introduces a dependence on the second QD transition. In the experiment, the QD and the cavity-mode splittings are fixed; this means that we can bring one of each in resonance but never the four simultaneously. Above, we already assumed resonance with the H-mode, $t_\text{H} = 1$. Fixing $\delta_{\rm cav}$ and $\delta_{\rm QD}$ to the experimental values, maximum extinction is achieved for $\Delta \omega_a/(2\pi) \simeq - 0.4~\text{GHz}$; this showcases that the combined effect of the second QD transition and the second cavity-mode shifts the optimal extinction away from the resonance of the main QD transition.

For the general scenario captured by~Eq.~\eqref{eq:tTresult} and realised experimentally, in which $\theta = 25.1^\circ, \beta = 0.92$, the maximum extinction in resonance with the H-mode is achieved for $\Delta \omega_a/(2\pi) = -0.31~\text{GHz}$, as shown in Fig.~\textbf{1}(d) of the main text. The colormap for the full theoretical model is displayed in Fig~\ref{fig:fig1_SI}(a) as a function of $\Delta\omega_a$ and $\Delta \omega_{\rm H}$; the white-dashed lines are the detunings at which the transition frequencies between the ground state and first few excited states of the free Hamiltonian $\hat{H}_0 + \hat{H}_{\rm int}$ [Eqs. (\ref{eq:ham0}, \ref{eq:hamint})] are zero (resonances).  

\subsection{Phase of the the transmitted light}

\begin{figure}[t]
    \centering
    \includegraphics[width = \textwidth]{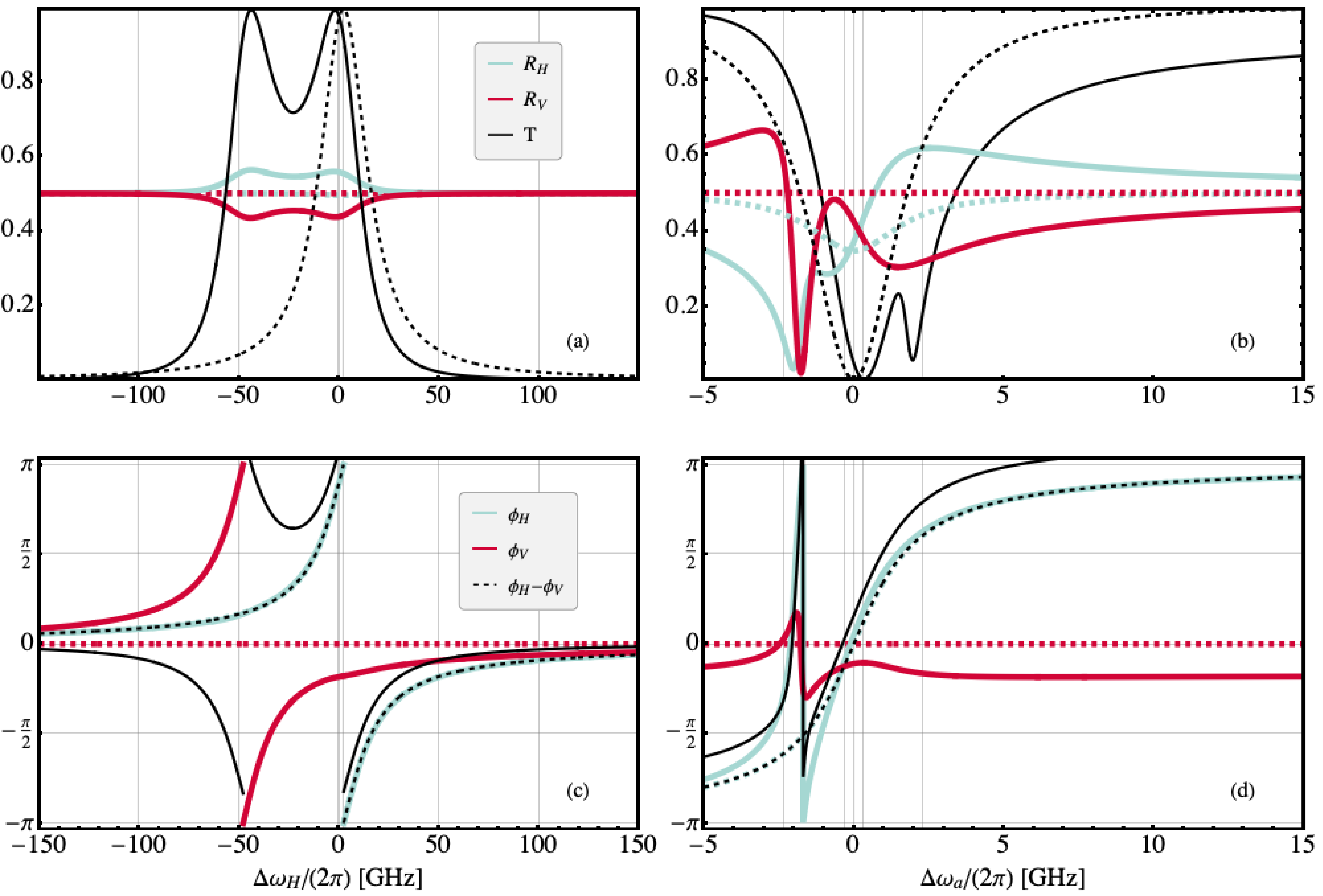} 
    \caption{
    Theoretical curves for the transmission and H/R reflection contrasted with the phase change. Mint-green and red correspond to H and R polarisations, respectively.
    Solid and dashed curves correspond to the full and JC models, respectively. For the phase plots, we use $\phi = \arg ( . )$, with the $\arg$-function defined in the range $[-\pi, \pi]$ (radians).  $\Delta \omega_x/(2\pi)$ are in units of GHz and $T/R's$ dimensional. For (a/c) we fix $\Delta\omega_a/(\pi) =10$ and for (b/d) $\Delta \omega_{\rm H}/(2\pi) = 0$.
    \textbf{(a)} The black solid (dashed) curve represents the transmission of the full (JC) model, in the far left regions of Figs.~\ref{fig:fig1_SI} (a/b), respectively. We plot separately only the reflection functions of H (mint-green) and V (red). In this plot, the contribution of the QD is negligible since it is far detuned.
    \textbf{(b) }With the same conventions of (a), we plot the transmission/reflection for $\Delta \omega_{\rm H} =0 $ sweeping the QD resonances [white dashed line of Fig.~\ref{fig:fig1_SI} (a/b) for full/JC models].
    \textbf{(c)} For the same parameters as (a), interference can be understood through the phases defined by Eqs.~(\ref{eq:phase1},~\ref{eq:phase2}). For the JC model, the relative phase, $\phi_H-\phi_V$, winds around $2\pi$ at $\Delta \omega_\text{H} = 0$, due to $\phi_{\rm H}$ (dashed mint-green and solid mint-green curves overlap), while $\phi_{\rm V}$ does not change. This explains the single peak structure in the dashed curve of (a) (M-light). For the full model, a similar behaviour is seen as both $\phi_{\rm V}$ and $\phi_{\rm H}$ wind around $2\pi$ across their respective resonances. The black solid curve thus showcases the relative phase, which is particularly relevant in the region between cavity resonances. This explains the double peak structure in (a).
    \textbf{(d)} Same parameters as in (b). For the JC model (dashed), $\phi_{\rm H}$ winds around $2\pi$ as the resonance $\Delta\omega_a = 0$ is crossed. This explains the single dip of the M-polarisation observed in (b) and the JC model for the maximal extinction discussed in the main text. For the full model (black-solid) the two QD transitions result in two $2\pi$ windings of the phase giving rise to the double-dip structure in (b) and in the experimental data in Fig. 1 (c) of the main text. The asymmetry in the dips in (b) is due to the $\ket{g} \leftrightarrow \ket{a}$ transition coupling mostly to the H-polarisation, which is on resonance.}
    \label{fig:figphase_SI} 
\end{figure}

From the definition of the transmitted amplitude \eqref{eq:tT} and Eqs.~\eqref{eq:IORsTM}, we can generally write the transmitted amplitude in terms of reflection amplitudes of H and V modes
\begin{align}\label{eq:phase1}
    t = \frac{|r_{\rm H}|}{\sqrt{2}}e^{i\phi_\text{H}} - \frac{|r_{\rm V}|}{\sqrt{2}}e^{i\phi_\text{V}},
\end{align}
where we introduced $r_\mu = \E{\bout{\mu}}/\E{\bin{\text{P}}},~\mu = \text{H, V}$ and wrote them in polar form, with $\phi_\mu = \arg(r_\mu)$. We emphasise that, for the transmission function, only the relative phase contributes
\begin{align}\label{eq:phase2}
    T = \frac{|r_\text{H}|^2}{2} + \frac{|r_\text{V}|^2}{2} - |r_\text{H} r_\text{V}|\cos(\phi_{\rm H} - \phi_{\rm V}),
\end{align}
and that $\phi_\mu$'s are defined relative to the absolute phase of the P-polarised input, which was fixed as $\arg(-i) = -\pi/2$ from the start.

In the JC model, light with V polarization cannot enter the cavity (since $|\delta_{\text cav}|\rightarrow \infty$) which implies that $|r_{\text V}|^2=1/2$ and $\phi_{\rm V}=0$, since only half the amount of photons are in the V-mode. Furthermore, if dissipation may be neglected, all the light entering the cavity eventually needs to be reflected, such that $|r_{\text H}|^2\simeq 1/2$. The transmission then simplifies to
\begin{align}\label{eq:phase3}
    T = \frac{1}{2}\left[1-\cos(\phi_{\rm H})\right].
\end{align}
Far away from any resonances, the H-mode light is reflected without a phase shift ($\phi_{\rm H}=0$) which implies $T=0$. When the incoming light crosses a resonance (e.g., at $\Delta\omega_{\rm H}=0$), $\phi_{\rm H}$ winds around $2\pi$ which results in a peak in the transmission located at the resonance, see Fig.~\ref{fig:figphase_SI}\,(a) and (c). When another resonance is crossed (e.g., at $\Delta\omega_{\rm a}=0$) while already on resonance with $\Delta\omega_{\rm H}=0$, the phase winds around $2\pi$ again, now resulting in a dip in transmission such that $T=0$ for $\Delta\omega_{\rm H}=\Delta\omega_{\rm a}=0$, see Fig.~\ref{fig:figphase_SI}\,(b) and (d). While this phase consideration captures many of the qualitative features observed in the experiment, the quantitative values for the transmission are complicated by dissipation, as well as the interplay between the two cavity modes and the two QD transitions.


\subsection{Saturation power}
As discussed in~\ref{ss:transmission}, the dip in transmission representing the maximum extinction in Fig. 1 (d) of the main text can be fully described by the first non-zero order in perturbation theory in $\epsilon$. For increasing photon-flux (or, equivalently, laser power), this dip is lifted  and can no longer be captured by a perturbative approach.

\begin{figure}[H]
    \centering
    \includegraphics[width = \textwidth]{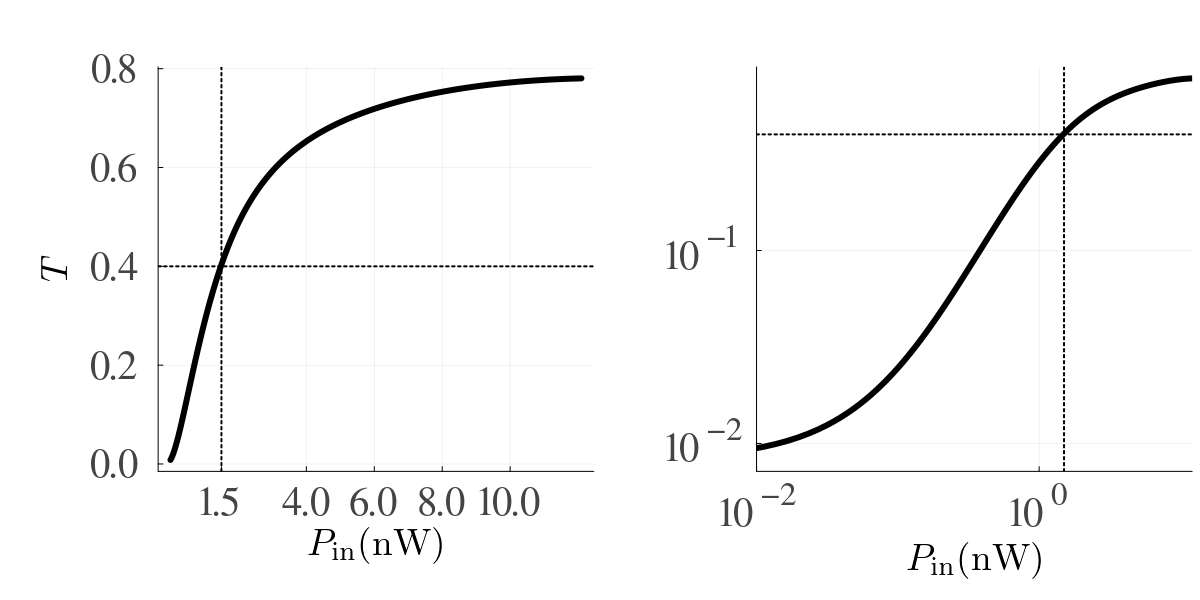} 
    \caption{Transmission as a function of $P_{\rm in}$ at maximum extinction. In GHz, we set $g/(2\pi) = 4.8,~ \kappa/(2\pi)=28,~\delta_\text{QD}/(2\pi)=2.3,~\delta_\text{cav}/(2\pi)=50, \gamma/(2\pi)=0.3$, $\theta = 25.1^\circ$ and $\Delta\omega_a/(2\pi) = - 0.31$ and $\omega_{\rm laser}/(2\pi) = 325$ THz. On the right, the same plot is produced in log-log scale, as in the inset of Fig. 1(d) in the main-text.}
    \label{fig:saturation_SI} 
\end{figure}

In order to investigate the power dependence of maximum extinction, in this subsection we present exact numerics with Hilbert space truncation in the cavity modes of $10$ photons. As mentioned in the main text, the laser power is measured just before entering the cavity, and corresponds to the \textit{input-power}
\begin{align}
    P_{\rm in} = \hbar \omega_{\rm laser} |\E{\bin{\rm P}}|^2 = \hbar \omega_{\rm laser} \epsilon^2~[\rm nW],
\end{align}
where we see that it is uniquely determined by $\epsilon$ and re-introduced $\hbar$ to highlight the conversion to nanowatts. In Fig.~\ref{fig:saturation_SI} we plot the transmission, which for high power saturates at $T=0.8$. The theory thus predicts a saturation power at half-saturation of transmission ($T=0.4$) as $P_{\rm sat}\approx 1.5$ nW.

In the main text we mention the experimental value of $1.8$ nW. This discrepancy may arise from losses which are not taken into account in the theory.

\subsection{Rice-Carmichael state}

Following Ref.~\cite{rice_single-atom_1988}, we now construct the reduced state of the M-cavity mode, $\hat{\rho}_\text{M}$. We start with the following trivial identities
\begin{align}\label{eq:Cide}
    \E{(\hat{a}^\dagger_\text{M})^m(\hat{a}_\text{M})^n}&=\sum_k \bra{k}_\text{M} \hat{a}^n_\text{M}  \hat{\rho}_\text{M}(\hat{a}^\dagger_\text{M} )^m\ket{k}_\text{M}\\
    &=\sum_k \frac{1}{k!}\bra{0}_\text{M} \hat{a}^{n+k}_\text{M}  \hat{\rho}_\text{M}(\hat{a}^\dagger_\text{M} )^{m+k}\ket{0}_\text{M}\nonumber\\
    &=\sum_k \frac{\sqrt{n!m!}}{k!}\bra{n+k}_\text{M} \hat{\rho}_\text{M}\ket{m+k}_\text{M},\nonumber
\end{align}
and we note that within the adiabatic elimination~\eqref{eq:adb}
\begin{align}\label{eq:aMadb}
    \hat{a}_\text{M}  = i\frac{\epsilon}{\sqrt{\kappa}}(t_\text{H}-t_\text{V})\qty(1- \frac{\sqrt{\Gamma}}{\epsilon (t_\text{H}-t_\text{V})}\hat{\sigma}_\text{M}),
\end{align}
where we introduced the QD operator which couples to the M polarisation of the cavity field
\begin{align}
    \hat{\sigma}_\text{M} &= \dyad{g}{\text{M}_\text{QD}}\\
    \bra{\text{M}_\text{QD}}&=\frac{(\tH \cos\theta + \tV\sin\theta)\bra{a}-(\tV \cos\theta + \tH\sin\theta)\bra{b}}{\sqrt{|\tH|^2+ |\tV|^2 + 2\Re{\tH \tV^*} \sin2\theta}}.
    \end{align}
According to Eq.~\eqref{eq:aMadb}, we must have $\expval{(\hat{a}_\text{M}^\dagger)^m \hat{a}_\text{M}^n}\propto  (\epsilon/\sqrt{\kappa})^{n+m}$ to leading order in $\epsilon/\sqrt{\kappa}$. Setting $k=0$ in Eq.~\eqref{eq:Cide} then self-consistently ensures that we always keep the leading order terms in $\epsilon/\sqrt{\kappa}$. This leads to a relation between the matrix elements of the reduced state of the cavity mode M and averages of the operators of the three-level system,
\begin{align}
\bra{n}_\text{M}\hat{\rho}_\text{M}\ket{m}_\text{M}=\frac{\E{(\hat{a}^\dagger_\text{M})^m(\hat{a}_\text{M})^n}}{\sqrt{n!m!}},
\end{align}
where the RHS of the above equation is computed from evaluating the effective QD averages given by Eq.~\eqref{eq:aMadb}, see Ref.~\cite{rice_single-atom_1988}. In this manner, we can construct the density matrix
\begin{align}
 \hat{\rho}_\text{M}= \dyad{\alpha} -\frac{\sqrt{\Gamma}}{\epsilon}\qty(\frac{\hat{N}_\text{M}\E{\hat{\sigma}_\text{M}}\dyad{\alpha}}{t_\text{H} - t_\text{V}} + \frac{\dyad{\alpha}\E{\hat{\sigma}_\text{M}}\hat{N}_\text{M}}{{t_\text{H}^* - t_\text{V}}^*}) + \frac{\Gamma}{\epsilon^2} \frac{\E{\hat{\sigma}^\dagger \hat{\sigma}}\hat{N}_\text{M}\dyad{\alpha}\hat{N}_\text{M}}{\qty|{t_\text{H} - t_\text{V}}|^2},
\end{align}
where $\ket{\alpha}$ is a coherent state and $\alpha = i (t_\text{H}-t_\text{V})\epsilon/\sqrt{\kappa}$. So far we only considered the adiabatic elimination (where $\epsilon/\sqrt{\kappa}$ has to be small); if we further consider the weak drive limit (where $\epsilon^2$ has to be small compared to all other energy scales), it holds  $\E{\hat{\sigma}^\dagger_{\rm M} \hat{\sigma}_{\rm M}}=|\E{\hat{\sigma}_{\rm M}}|^2$ and the reduced state is a pure state~\cite{Carmichael2008-cQED2}, $\hat{\rho}_\text{M}=\dyad{\psi}$
\begin{align}
    \ket{\psi} = \qty(1-\frac{\sqrt{\Gamma}}{\epsilon}\frac{\E{\hat{\sigma}_\text{M}}}{{t_\text{H} - t_\text{V}}}\hat{N}_\text{M})\ket{\alpha}.
\end{align}
The above structure for the reduced state of the M mode describes not only the steady-state, but also its evolution after photo-detection. This is true due to the factorization of QD transition operators~$\expval{\sigma_i^\dagger \sigma_j}_\tau$, as discussed in~Ref.~\cite{Carmichael2008-cQED2}. Due to this property, we can write the Rice-Carmichael state at arbitrary times after photo-detection as
\begin{align}\label{eq:TMCarmichael}
    \ket{\psi}_\tau &= \qty(1-\sqrt{\frac{\Gamma}{\epsilon^2}}\frac{1}{(\tH-\tV)}\E{\hat{\sigma}_\text{M}}_\tau\hat{N}_\text{M})\ket{\alpha}\\
                                &= \qty(1-\sqrt{\frac{\Gamma}{\kappa}}\E{\hat{\sigma}_\text{M}}_\tau\hat{a}_\text{M}^\dagger)\ket{\alpha},\nonumber
\end{align}
where
\begin{align}\label{eq:dipoleM}
    \E{\hat{\sigma}_\text{M}}_\tau=\tr[ \hat{\sigma}_\text{M}(\tau)\frac{\hat{a}_\text{M}(0)\hat{\rho}_\sss\hat{a}_\text{M}^\dagger(0)}{\E{\hat{a}^\dagger_\text{M}\hat{a}_\text{M}}_\infty}],
\end{align}
 can be evaluated from the perturbative propagator~\eqref{eq:dyson} and using the adiabatic elimination \eqref{eq:aMadb}. Equation~\eqref{eq:dipoleM} is an average conditioned on a photo detection. In the JC model ($\theta =0 $ and $t_\text{V}\to 0$), the Rice-Carmichael assumes a simple form
\begin{align}\label{eq:twosidedcstate}
   \ket{\psi}_\tau = \qty[1-\frac{F_\text{P}t_\text{H}}{1+ F_\text{P}t_\text{H}}\qty(1+ F_\text{P}t_\text{H} e^{-\frac{\gamma \tau}{2}(1+F_\text{P}t_\text{H})})\hat{N}_\text{M}]\ket{\alpha}, 
\end{align}
Equation (2) of the main text is obtained on cavity resonance, $t_\text{H}=1$, writing $F_\text{P}= \beta/(1-\beta)$ and evaluating the average of $\hat{a}_{\rm M}$.

\begin{figure}[t]
    \centering
    \includegraphics[width = \textwidth]{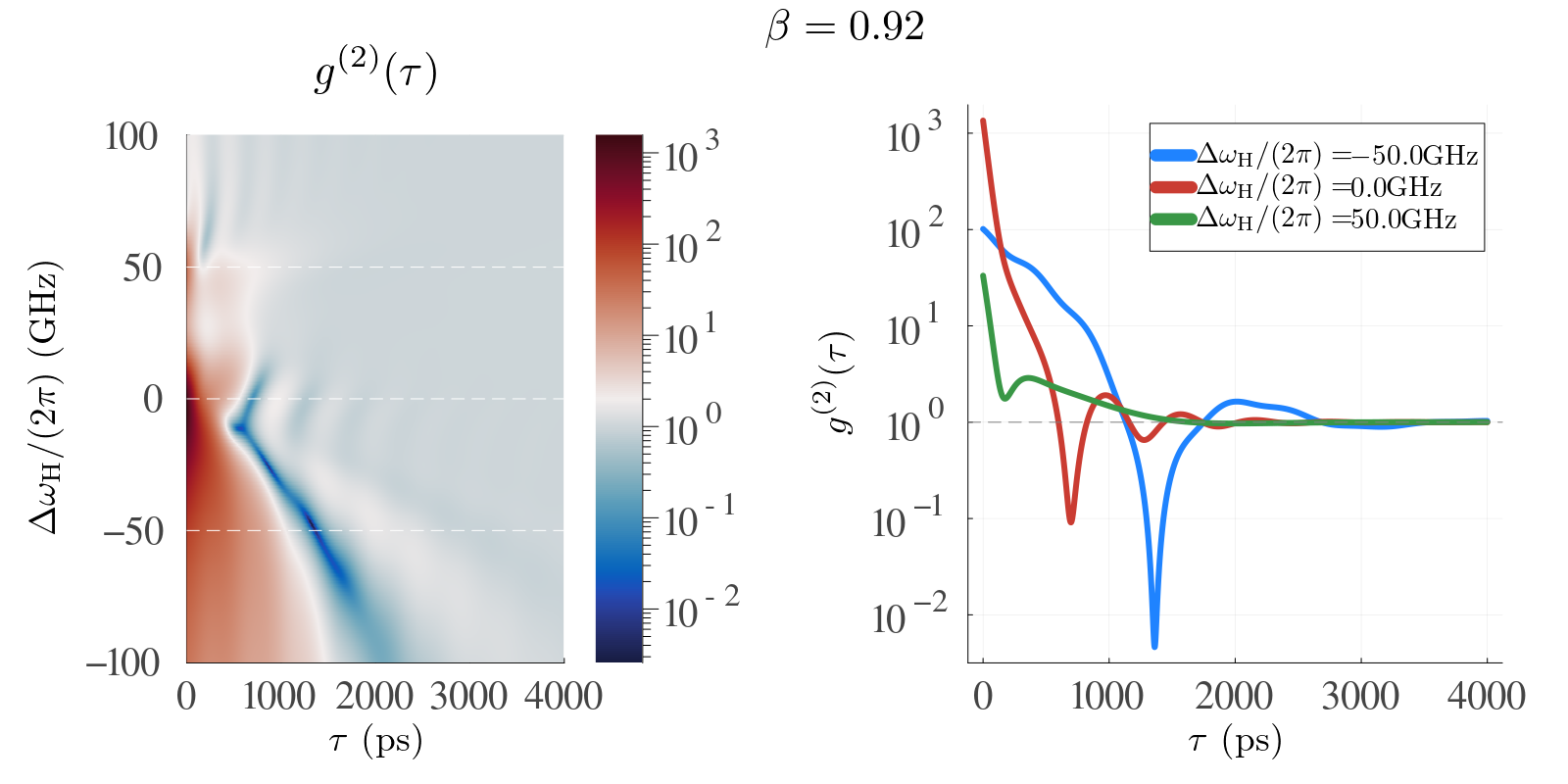} 
    \caption{Theoretical results of transmission mode intensity correlations. In GHz, we set $g/(2\pi) = 4.8,~ \kappa/(2\pi)=28,~\delta_\text{QD}/(2\pi)=2.3,~\delta_\text{cav}/(2\pi)=50, \gamma/(2\pi)=0.3$ and $\theta = 25.1^\circ$ corresponding to $\beta=0.92$. The white dashed lines correspond to the cuts displayed in the plot; the red line is the H-cavity resonance (discussed in the main text), the blue line is the V-cavity resonance. We also plot $\Delta\omega_H=\delta_{\rm cav}$, green line .}
    \label{fig:fig2_SI}
\end{figure}

For the full model, representing the experimental scenario, we show the plots of the $g^{(2)}(\tau)$ in different cavity detuning regimes and high $\beta$ in Fig.~\ref{fig:fig2_SI}. In these plots we also consider $\Delta \omega_a/(2\pi) = -0.14 \text{GHz} $; this value is not the one at optimum extinction predicted in the main text, but the difference between this value lies within the precision of the gate voltage control of the QD. 

\subsection{Intensity correlations}

The Rice-Carmichael can be used to compute the intensity correlations and also to interpret these measurements.  The connection between Eq.~\eqref{eq:TMCarmichael} and the $g^{(2)}(\tau)$-- function measured in the experiment is
\begin{align}
    g^{(2)}(\tau) = \qty|\frac{\E{\hat{a}_\text{M}}_\tau}{\E{\hat{a}_\text{M}}_\infty}|^{2},
\end{align}
where the average is taken w.r.t. the RC-state~\eqref{eq:TMCarmichael}. This highlights that for a generic RC-state of the form
\begin{align}
    \ket{\psi}_\tau = (1-C_\tau \hat{N}) \ket{\alpha},
\end{align}
where we drop the subscript $\text{M}$ for simplicity. We may write $g^{(2)}(\tau)= |\mathcal{A}(\tau)|^2$ where the amplitude $\mathcal{A}(\tau)=\E{a}_\tau/\E{a}_\sss$ can be computed solely from the Rice-Carmichael
\begin{align}
    \langle\psi|\hat{a}|\psi\rangle_\tau = \alpha(1-C_\tau) + \mathcal{O}(\alpha^3),
\end{align}
which gives
\begin{align}
    \mathcal{A}(\tau) = \frac{1 - C_\tau}{1-C_\infty}.
\end{align}

For the full experimental scenario, the explicit expression is large and used only for numerics. We now contrast the experimental scenario to limiting scenarios mentioned in the main text. According to Eq.~\eqref{eq:twosidedcstate}, we obtain for the JC model
\begin{align}
    g^{(2)}(\tau) = \qty|1-(F_\text{P} t_\text{H})^2e^{-\frac{\gamma \tau}{2}(1+ F_\text{P} t_\text{H})}|^2,
\end{align}
which is precisely the same expression as for the transmission in a two-sided cavity. For H-cavity resonance, $\tH = 1$, we obtain the same expression as derived in~Ref.~\cite{rice_single-atom_1988} and discussed in the main text
\begin{align}
    g^{(2)}(\tau) = \qty|1-\frac{\beta^2}{(1-\beta)^2}e^{-\frac{\gamma \tau}{2(1-\beta)}}|^2,
\end{align}
which is written in terms of $\beta = F_\text{P}/(F_\text{P}+1)$. For $\beta=1$, $g^{(2)}(0)\to\infty$, indicating that only multi-photon, $\ket{n>1}$, components are transmitted, a highly bunched state.
Notably, for $1/2 \leq \beta \leq 1$ there exists a time $\tau_0(\beta)$ such that $g^{(2)}(\tau_0) = 0$
\begin{align}
    \tau_0 = \frac{2}{\gamma}(1-\beta) \ln\qty[\frac{\beta^2}{(1-\beta)^2}].
\end{align}

For the complete experimental scenario, Fig. \ref{fig:fig2_SI} displays intensity correlations at high-$\beta$ and its dependence on the H-cavity detuning. 

\subsection{Dephasing analysis}
\begin{figure}[h]
    \centering
    \includegraphics[width = .5\textwidth]{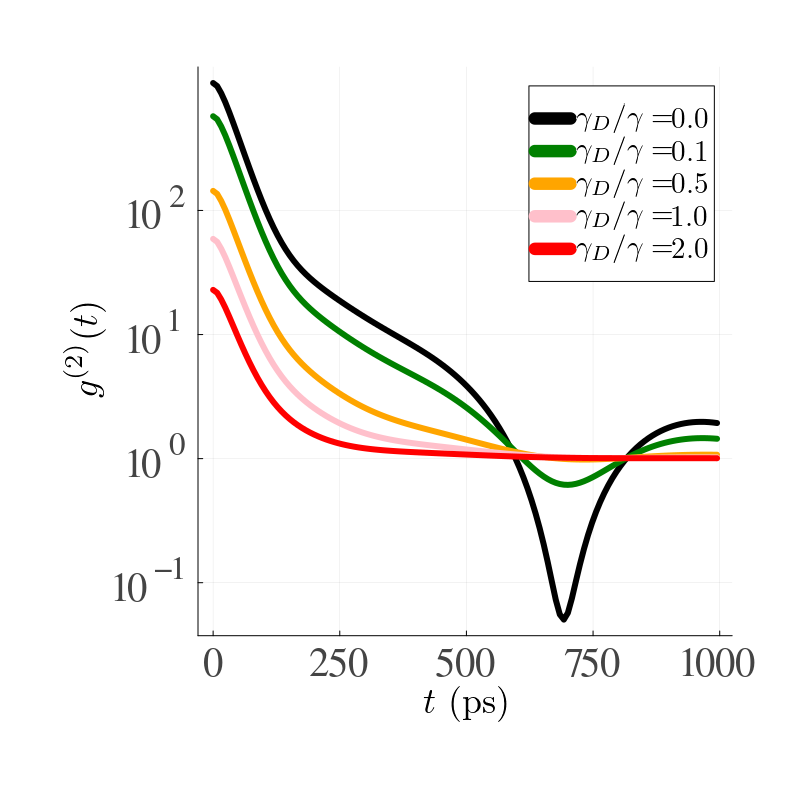} 
    \caption{transmission mode intensity correlations in the presence of dephasing. In GHz, we set $g/(2\pi) = 4.8,~ \kappa/(2\pi)=28,~\delta_\text{QD}/(2\pi)=2.3,~\delta_\text{cav}/(2\pi)=50, \gamma/(2\pi)=0.3$, $\Delta\omega_{\rm H} = 0$ and $\theta = 25.1^\circ$, corresponding to $\beta=0.92$.}
    \label{fig:dephasing_SI}
\end{figure}
In order to confirm the robustness of this cavity-QED setup against dephasing, we add to the LME~\eqref{eq:lme} the dissipator,
\begin{align}
\mathcal{D}_{\gamma_D} \hat{\rho} = \gamma_D\qty(D[\hat{\sigma}_a^\dagger \hat{\sigma}_a]\hat{\rho} + D[\hat{\sigma}_b^\dagger\hat{\sigma}_b]\hat{\rho}),
\end{align}
and simulate the $g^{(2)}$--function for different values of $\gamma_D/\gamma$. Figure~\ref{fig:dephasing_SI}  highlights that dephasing has the effect of reducing the bunching at zero delay and the anti-bunching at finite delay. Due to the strong agreement of the experiment with the theory without the dephasing and that dephasing would destroy these coherent features, we verify that $\gamma_D$ is negligible.

\section{Reflection mode experiment}
\subsection{Reflection}
Under the adiabatic elimination and weak drive approximations, we now provide the analytical expression for Eq.~\eqref{eq:rBR}
\begin{align}\label{eq:rBRresult}
    r_\leftarrow = 1- 2 t_\text{H} +2 \frac{\beta \mathcal{N}_1 +\frac{\beta^2}{2}\mathcal{N}_2}{(\gamma - 2i\Delta\omega_a)(\gamma - 2i\Delta\omega_b) + \beta \mathcal{D}_1 + \frac{\beta^2}{2}\mathcal{D}_2},
\end{align}
where the coefficients are,
\begin{align}
    \mathcal{N}_1 =& t_\text{H}^2 \gamma (\gamma- i(\Delta\omega_a - \Delta \omega_b)(1-\cos2\theta)), \\
    \mathcal{N}_2 =& t_\text{H}^2 \gamma((-2+ t_\text{V})\gamma+ 2i(\Delta \omega_a + \Delta \omega_b) + 2i(\Delta \omega_a -\Delta \omega_b)\cos 2\theta + \gamma t_\text{V} \cos 4\theta),\\
    \mathcal{D}_1 =&(t_\text{H} + t_\text{V} -2 )\gamma^2 + 8 \Delta \omega_a \Delta \omega_b - i\gamma(t_\text{H} + t_\text{V} - 4)(\Delta \omega_a +\Delta \omega_b)+ i\gamma(t_\text{H} - t_\text{V})(\Delta \omega_a -\Delta \omega_b) \cos 2\theta),\\
    \mathcal{D}_2 =&(2-2t_\text{H}-2t_\text{V} + t_\text{H} t_\text{V})\gamma^2 - 8 \Delta \omega_a \Delta \omega_b + 2i (t_\text{H} + t_\text{V} -2 )(\Delta \omega_a + \Delta \omega_b)\gamma \\ 
    &- 2i\gamma (t_\text{H} - t_\text{V})(\Delta \omega_a - \Delta \omega_b)\cos 2\theta + \gamma^2 t_\text{H} t_\text{V} \cos 4 \theta.  \nonumber 
\end{align}
For this scenario, most of the physics can be understood by looking at the JC model, obtained in the limiting case of $\theta = 0$, in which there is no way light can populate the V polarisation: The light scatters only in the H port (or is lost through dissipation). This means that for $\theta =0$, $t_\leftarrow = 0$ independently of any other parameter choice and we find
\begin{equation}
    r_\leftarrow = 1 - 2t_\text{H}\left[1-  \frac{2 t_\text{H}\beta }{1-(1-t_\text{H})\beta - 2i(1-\beta)\Delta \omega_a/
\gamma}\right],
\end{equation}
 at $\Delta \omega_a=0$, the reflection probability is thus $R_\leftarrow = (1 - 2\beta)^2$. We now consider a perturbation in $\theta$, in Eq.~\eqref{eq:rBRresult}, and, for simplicity, $\Delta \omega_\text{H} = 0$
 \begin{align}
     r_\leftarrow = -1 &+ \frac{2\beta \gamma}{\gamma - 2i(1-\beta)\Delta \omega_a}\\
      &- \theta^2\frac{4(1-\beta)^2\beta \gamma(\gamma - 2i\Delta \omega_a)(\Delta \omega_a - \Delta \omega_b)}{(\gamma - 2i(1-\beta)\Delta \omega_a)^2(\gamma + (t_\text{V} - 1)\beta \gamma - 2i\Delta \omega_b(1-\beta))}\nonumber\\
     &+\theta^2\frac{ 4t_\text{V}(1-\beta)\beta \gamma (2\gamma - i(3\Delta \omega_a + \Delta \omega_b))}{(\gamma - 2i(1-\beta)\Delta \omega_a)^2(\gamma + (t_\text{V} - 1)\beta \gamma - 2i\Delta \omega_b(1-\beta))} .\nonumber
 \end{align}
 Above, the second line showcases a correction coming from the second QD transition, which now also couples to the H-mode and disappears if the splitting is zero, $\Delta \omega_a = \Delta \omega_b$. The third line highlights the correction coming from the coupling between QD transition $a$ and the V-mode, which can be discarded if V is far off-resonant, $t_\text{V} \to 0$. We also note that both pertubartive corrections can be suppressed by a high $\beta$-factor, $\beta\simeq 1$.

\begin{figure}[t]
    \centering
    \includegraphics[width = \textwidth]{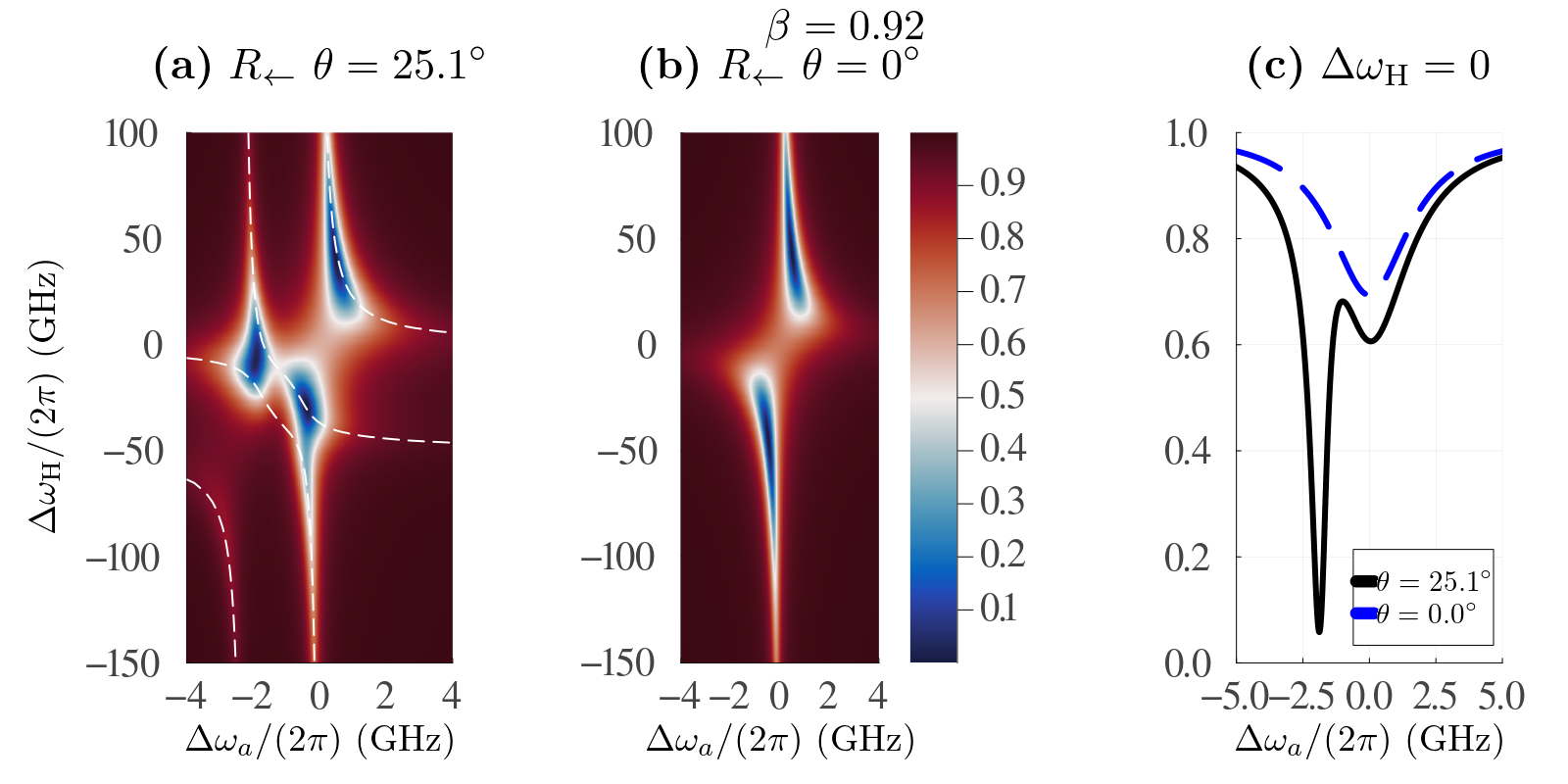} 
    \caption{Theoretical results for reflection, to be compared with Fig.~3\,(a) in the main text. In GHz, we set $g/(2\pi) = 4.8,~ \kappa/(2\pi)=28,~\delta_\text{QD}/(2\pi)=2.3,~\delta_\text{cav}/(2\pi)=50, \gamma/(2\pi)=0.3$. \textbf{(a)} Here we take $\theta = 25.1^\circ$. The black lines are the points in which the gap between the ground and the first four excited states of $\hat{H}_0 + \hat{H}_{\rm int}$ (Eqs.\eqref{eq:ham0} and \eqref{eq:hamint}) close. \textbf{(b)} For comparison, we consider $\theta=0$ (JC model).~\textbf{(c)} Reflection function on cavity resonance $\Delta \omega_\text{H}=0$ for both $\theta = 25.1^\circ$ and $\theta =0$.}
    \label{fig:fig3_SI}
\end{figure}

 In the colormap, ~Fig.~\ref{fig:fig3_SI}(a), we plot the full theoretical model as a function of $\Delta \omega_a$ and $\Delta \omega_{\rm H}$; the white-dashed lines are the same as explained in the last paragraph of Subsection~\ref{ss:transmission}.
\subsection{Rice-Carmichael state}
For this configuration, the output port is coupled to the H-cavity mode. Accordingly, we construct the Rice-Carmichael of the H-cavity mode, following a similar procedure to the transmission mode, we use Eqs.~\eqref{eq:adb} to construct 
\begin{align}\label{eq:BRCarmichael}
    \ket{\psi_\leftarrow} = \qty( 1-\sqrt{\frac{\Gamma}{\epsilon^2}} \E{\hat{\sigma}_\text{H}}_\tau\hat{N}_\text{H})\ket{\alpha},
\end{align}
where $\alpha = 2i\tH \epsilon/\sqrt{\kappa}$,
\begin{align}
\hat{\sigma}_\text{H} = \Sa \cos\theta - \Sb \sin \theta,
\end{align}
and the time-dependent average now must take into account the laser background through the displaced operator $ \hat{b}_\text{H}=\hat{a}_\text{H}-i\epsilon/\sqrt{\kappa}$
\begin{align}
    \E{\hat{\sigma}_\text{H}}_\tau=\tr[ \hat{\sigma}_\text{H}(\tau)\frac{\hat{b}_\text{H}(0)\hat{\rho}_\sss\hat{b}_\text{H}^\dagger(0)}{\E{\hat{b}^\dagger_\text{H}\hat{b}_\text{H}}_\infty}].
\end{align}
We note that the operator $\hat{b}_\text{H}$ is introduced in agreement with the IOR~\eqref{eq:ior}, and can be used to rewrite normal-ordered correlation functions of the reflected output field in terms of cavity operators [see, for instance, the formal expression for $g^{(2)}_\leftarrow(\tau)$~\eqref{eq:formalg2BR}].


\subsection{Intensity correlations}
\begin{figure}[t]
    \centering
    \includegraphics[width = \textwidth]{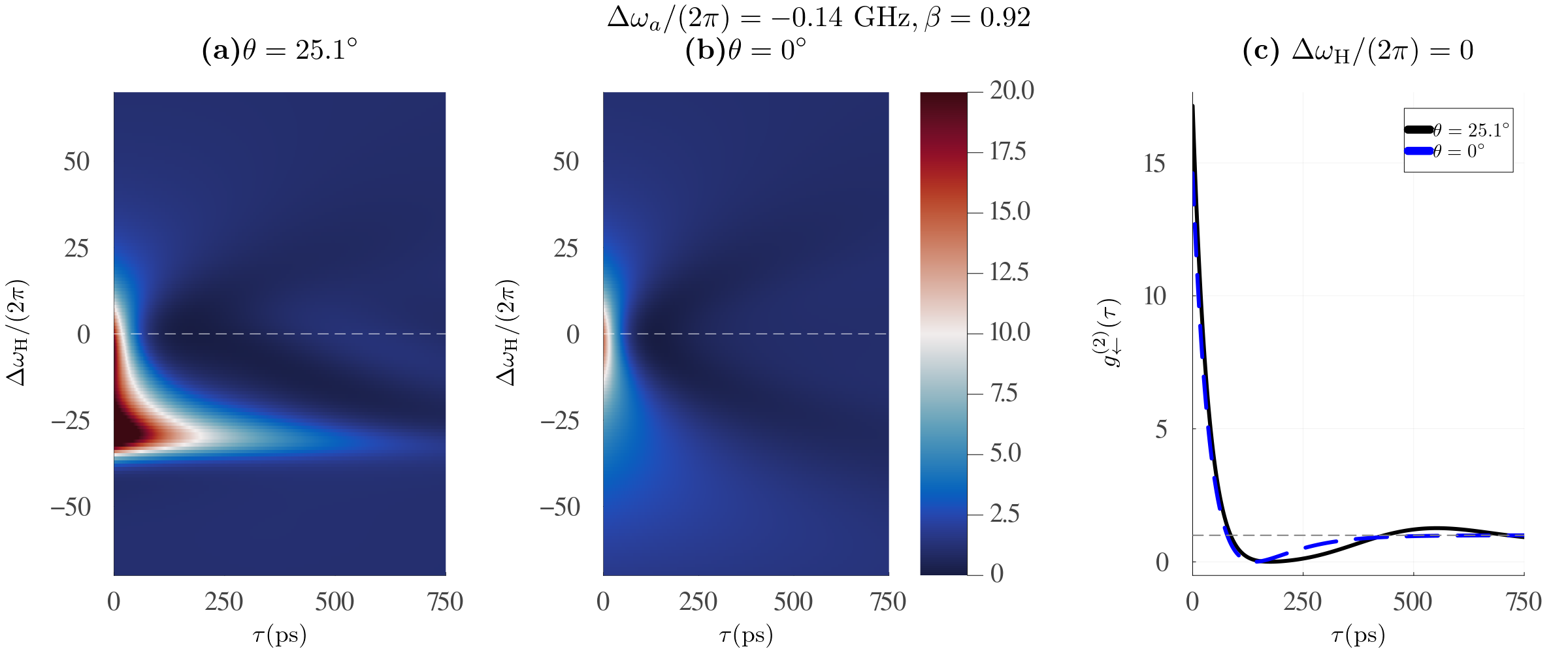} 
    \caption{Theoretical results of reflection intensity correlations. In GHz, we set $g/(2\pi) = 4.8,~ \kappa/(2\pi)=28,~\delta_\text{QD}/(2\pi)=2.3,~\delta_\text{cav}/(2\pi)=50, \gamma/(2\pi)=0.3$ and. $\Delta \omega_a \neq 0 $ due to the drift of the QD resonance during the measurements. (a) Full model, corresponding to the experimental scenario, $\theta = 25.1^\circ$. (b) JC model, equivalent to a two-level system in a single mode, one-sided cavity, $\theta = 0^\circ$.  (c) Line cuts of the cavity resonance, $\Delta \omega_{\rm H} =0$ for both full model and JC model.}
    \label{fig:fig4_SI}
\end{figure}
The intensity correlations are fully determined by evaluating the following average w.r.t. the Rice-Carmichael~\eqref{eq:BRCarmichael}
\begin{align}
g^{(2)}_\leftarrow (\tau) = \qty|\frac{\E{ \hat{b}_\text{H}}_\tau}{\E{\hat{b}_\text{H}}_\infty}|^2.
\end{align}
We can then write $g^{(2)}_\leftarrow(\tau)=|\mathcal{A}_\leftarrow(\tau)|^2$ in terms of $C_\tau = \sqrt{\Gamma/\epsilon^2}\E{\hat{\sigma}_\text{H}}_\tau$. 

For simplicity, we briefly concentrate on the resonant case $\tH =1 $, in which case $\hat{b} = \hat{a} - i\epsilon/\sqrt{\kappa} = \hat{a} - \alpha/2$ (dropping the subscript \text{H}). Thus
\begin{align}
    \bra{\psi_\leftarrow}\hat{b}\ket{\psi_\leftarrow}_\tau = \frac{
    \alpha}{2} \qty(1- 2 C_\tau) + \mathcal{O}
(\alpha^3),
\end{align}
and
\begin{align}
    \mathcal{A}_\leftarrow(\tau) = \frac{1- 2C_\tau}{1-2C_\infty}.
\end{align}
For $\theta =0$ we obtain the amplitude
\begin{align}
    \mathcal{A}_\leftarrow(\tau) =1- \qty(\frac{2\beta}{1-2\beta})^2 e^{-\frac{\gamma \tau}{2(1-\beta)}},
\end{align}
resulting in
\begin{align}
     g^{(2)}_\leftarrow(\tau) = \qty[1- \qty(\frac{2\beta}{1-2\beta})^2 e^{-\frac{\gamma \tau}{2(1-\beta)}}]^2,
\end{align}
where we used $\beta  = F_\text{P}/(1+F_\text{P})$. Notably, for $1/4 < \beta \leq 1$ there exists a time $\tau_0$ such that $g^{(2)}(\tau_0) = 0$
\begin{align}
    \tau_0 = \frac{2}{\gamma}(1-\beta) \ln\qty[\frac{(2\beta)^2}{(1-2\beta)^2}].
\end{align}

Off cavity resonance, we can still obtain simple results for $\theta=0$
\begin{align}\label{eq:g2brsimpl}
    g^{(2)}_\leftarrow(\tau) = \qty|1 -  4 F_\text{P}^2t_\text{H}^4\frac{\exp{-(\tau \gamma) /2 \qty[(1+ F_\text{P} \Re t_\text{H})+i\qty(2\Delta \omega_a/\gamma - F_\text{P} \Im t_\text{H})]}}{\qty[1- 2i\Delta \omega_a/\gamma -t_\text{H}((2-F_\text{P})-4i\Delta\omega_a/\gamma)]^2}|^2. 
\end{align}
In the above exponential, real and imaginary parts of $t_\text{H}$ dressed by the Purcell factor effectively reduce the Purcell-enhanced decay and introduce oscillations, respectively. Further, we note that $\Delta \omega_a/\gamma$ introduces oscillations.  In the denominator, we note that all these parameters contribute to $g^{(2)}(0)$ by shifting the maximum and minimum values it can attain.

For the experimental scenario and high-$\beta$ we plot the heatmap for the intensity correlations with respect to the H-cavity detuning in Fig.~\ref{fig:fig4_SI}. We can see that in this experiment we operate in a regime that is well captured by the JC model around $\Delta\omega_{\rm H}=0$. 
\newpage
\section{Parameters used in the figures of the main text}
\setlength{\arrayrulewidth}{0.5mm}
\setlength{\tabcolsep}{3pt}
\renewcommand{\arraystretch}{1.5}

In this Section we provide the parameters used in the plots of the main text, see Tab.~\ref{tab:my_label}. Parameters that are the same in all figures are (in GHz): $\kappa/(2\pi)=28$, $\delta_\text{QD}/(2\pi)=2.3$, $\Delta \omega_{\rm H}=0$ and $\gamma/(2\pi)=0.3$. All plots were obtained in the small drive limit, where the results do no longer change upon decreasing $\epsilon$. For the full model, we always used $\delta_\text{cav}/(2\pi)=50$ and $\theta=25.1^\circ$ whereas we used $\theta=0$ and $|\delta_\text{cav}|\rightarrow \infty$ for the JC model. Furthermore, for the JC model, we always used $\Delta \omega_a=0$ to operate at maximum extinction. For the full model, we employed different values in $\Delta \omega_a$ due to the limited experimental control on the gate voltage that determines this parameter. The theory predicts that at $\beta = 0.92$ the value of $g^{(2)}(0)$ is above $600$ at maximum extinction, $\Delta \omega_a/(2\pi) = -0.31$.


\begin{table}[h]
    \centering
    \begin{tabular}{|c||cc|}
    \hline
       ~  &  $g/(2\pi)$  &  $\Delta \omega_a/(2\pi)$, full model  \\ [0.5ex]
       \hline 
       \hline 
       Fig. 1(d)  & 4.8 GHz &  - \\
       Fig. 1(d) - inset  & 4.8 GHz  & -0.31 GHz\\
       Fig. 2(a)  & 4.8 GHz   & -0.31 GHz \\
       Fig. 2(b) - top left & 4.8 GHz  & -0.14 GHz \\
       Fig. 2(b) - top right   & 2.4 GHz & -0.14 GHz \\
       Fig. 2(b) - bottom right   & 1.4 GHz & -0.14 GHz \\
       Fig. 2(b) - bottom left   & 0.65 GHz & -0.14 GHz \\
       Fig. 3(a) - top & 4.8 GHz  & - \\
        Fig. 3(b) - top & 4.8 GHz & -0.14 GHz \\
       \hline
    
    \end{tabular}
    \caption{Parameters for theory plots in main text. Whenever a quantity features in the plot axes we mark it with ``-''. The JC model uses $\Delta \omega_a=0$ because this is where the extinction is maximal. In the full model, extinction is maximal at $\Delta \omega_a/(2\pi) = -0.31$. Deviations from this value are due to limited experimental control.}
    \label{tab:my_label}
\end{table}


\bibliography{supplement_references}